\documentclass[reprint, groupedaddress, showpacs, showkeys, aps]{revtex4-1}

\usepackage[T1]{fontenc}
\usepackage[utf8]{inputenc}
\usepackage{amsmath}
\usepackage{amssymb}
\usepackage[toc,page]{appendix}
\usepackage{graphicx} 
\usepackage{dcolumn}  
\usepackage{bm}       
\usepackage{float}

\newcommand{\be}{\begin{equation}}
\newcommand{\ee}{\end{equation}}
\newcommand{\bee}{\begin{equation*}}
\newcommand{\eee}{\end{equation*}}
\newcommand{\ud}{\mathrm{d}}

\newcommand{\norm}[1]{\left\lVert#1\right\rVert}
\newcommand{\prodscal}[2]{\left\langle#1,#2\right\rangle}

\begin{document}
\title{Periodic orbits in nonlinear wave equations on networks} 

\author{Jean-Guy CAPUTO}
\email{caputo@insa-rouen.fr}

\author{Imene KHAMES}
\email{imene.khames@insa-rouen.fr}

\author{Arnaud KNIPPEL}
\email{arnaud.knippel@insa-rouen.fr}

\affiliation{Laboratoire de Mathématiques, INSA de Rouen, \\ 76801 Saint-Etienne du Rouvray, France}

\author{Panayotis PANAYOTAROS}
\email{panos@mym.iimas.unam.mx}
\affiliation{Depto. Matemáticas y Mecánica, IIMAS-UNAM, Apdo. Postal 20-126,
01000 México DF, México}

\date{\today}

\begin{abstract}
We consider a cubic nonlinear wave equation on a network and show
that inspecting the normal modes of the graph, we can immediately
identify which ones extend into nonlinear periodic orbits. Two
main classes of nonlinear periodic orbits exist: modes without soft nodes
and others. For the former which are the Goldstone and the bivalent modes,
the linearized equations decouple.
A Floquet analysis was conducted systematically for
chains; it indicates that the Goldstone mode is usually stable and
the bivalent mode is always unstable. The linearized equations
for the second type of modes are coupled, they indicate which
modes will be excited when the orbit destabilizes. Numerical results
for the second class show that modes with a single eigenvalue are
unstable below a treshold amplitude. Conversely, modes with multiple
eigenvalues seem always unstable. This study could be applied to 
coupled mechanical systems. 
\end{abstract}

\maketitle
\section{Introduction}

Linearly coupled mechanical systems are well understood in
terms of normal modes, see \cite{landau}. These are bound states 
of the Hamiltonian which is a quadratic, symmetric function of positions and
velocities. The bound states are orthogonal and correspond to real
frequencies. Because of the orthogonality, normal modes do not couple
and the system can be described solely in terms of the amplitude of
each mode and its time derivative. When non linearity is present in
the equations of motion, normal modes will couple. Natural questions
are : how do they couple ? Is there any trace of them in the nonlinear
regime? An important realisation is that some systems exhibit nonlinear
periodic solutions. The situation differs whether the 
degrees of freedom are nonlinear and coupled linearly or whether they are linear oscillators
coupled nonlinearly as in the celebrated Fermi-Pasta-Ulam model \cite{fpu}.

Nonlinear oscillators coupled linearly can give rise to periodic solution
labeled "intrinsic localized modes", see the reviews \cite{flach98,flach08}. 
These are nonlinear periodic orbits that are exponentially
localized in a region of the lattice. For large amplitudes,
these solutions are in general different from the linear normal 
modes. Another type of nonlinear
periodic orbit can exist, which is a continuation of the linear normal 
modes \cite{panos}.
For the Fermi-Pasta-Ulam system, this type of
nonlinear periodic orbit has been found using group theoretical methods
and Hamiltonian perturbation methods \cite{cr12,crz05,bcs11}.
Also, in the theoretical mechanics community, such linear-nonlinear 
periodic orbits
have been studied for some time, see the extensive review by \cite{am13}.

In a pioneering work, for one dimensional lattices with linear coupling and
onsite nonlinearity, i.e. nonlinear oscillators coupled linearly,
Aoki \cite{aoki} recently found
families of nonlinear periodic orbits stemming from the linear normal modes.
He studied one dimensional lattices with periodic, fixed or free boundary conditions.
His main findings are that normal modes with coordinates containing $\pm 1$ and $0$ 
give rise to nonlinear periodic orbits. Analysis of their dynamical
stability revealed that the modes containing only $\pm 1$
lead to decoupled variational equations, one for each normal mode.

In this article, we extend Aoki's approach to general networks with linear couplings
and onsite nonlinearity. The underlying linear
model is the graph wave equation \cite{cks13} where the Laplacian is the graph
Laplacian \cite{Cvetkovic1}. It
is a natural description of miscible flows on a network since
it arises from conservation laws. It also models the density
when one considers a probabilistic motion on a graph.
In a recent work \cite{cks13}, we studied this model and showed
the importance of the normal modes, i.e. the eigenvectors of the Laplacian
matrix and their associated real eigenvalues.
We choose a cubic nonlinearity because the solution exists for all times, the
evolution problem is well-posed. 
We extend the criterion of Aoki to any network; by inspecting the normal
modes of a network we can immediately identify nonlinear periodic
orbits. We give all such nonlinear periodic orbits for cycles and chains,
these are the well-known Goldstone mode and what we call bivalent and trivalent modes.
We also show examples in networks that are neither chains or cycles.
Two main classes of modes exist, the ones with no zero 
nodes (the Goldstone and bivalent modes) and the others (trivalent modes). 
For the first class, the variational equations decouple completely into
$N$ Hill-like resonance equations. For the second
class, we give explicitely their form, enabling prediction of the 
couplings. It is surprising that the dynamics of periodic orbits is
different when soft nodes are present. The special role of these soft nodes
in the dynamics had been analyzed in \cite{cks13}.
For the first class we calculate the stability diagram of
the nonlinear periodic orbits systematically for chains. The Goldstone
mode is usually stable for large enough amplitude. On the contrary 
the bivalent modes are always unstable.
The second class is more difficult to study; it's stability is governed
by a system of coupled resonance equations. These reveal which new modes will
be excited when there is instability. Numerical calculations illustrate
these different situations. The fact that the nonlinear periodic 
orbits have an explicit solution, and the form of the linearization 
around some of these solutions are quite unique features of the model. 
We believe these results will be useful to the lattice community but also
more generally to the theoretical mechanics community where these systems occur.\\
The article is organized as follows:
We introduce the model in section 2. In section 3, we generalize the
criterion of Aoki that shows which linear normal modes extend
to nonlinear periodic orbits. Section 4 classifies these
nonlinear normal modes for chains and cycles and gives other examples.
Section 5 presents the variational equations obtained by perturbing
the nonlinear normal modes. These are solved numerically for chains
in section 6 for the Goldstone and the bivalent modes.
Full numerical results are presented in section 7 for trivalent modes.
Conclusions are presented in section 8.

\section{Amplitude equations}

We consider the following nonlinear wave equation on a connected graph with $N$ nodes
\begin{equation} \label{phi4}
\ddot{\mathbf{u}}=\mathbf{\Delta} \mathbf{u} - \mathbf{u}^3,
\end{equation}
where $\mathbf{u}=(u_1(t),\dots,u_N(t))^{T}$ is the field amplitude, 
$\mathbf{\Delta}$ is the graph Laplacian \cite{Cvetkovic1} with components $\Delta_{ij}, ~~1 \leq i,j \leq N $, 
$\mathbf{u}^3=(u_1^3,u_2^3,\dots,u_N^3)^{T}$ and where
$\ddot{\mathbf{u}} \equiv {\ud^2 {\mathbf{u}} \over \ud t^2}$.
Notice that we use bold-face capitals for matrices and bold-face lower-case letters for vectors.
This model is an extension to a graph of the simplified $\phi^4$ well-known model in condensed matter physics lattices \cite{Scott1}. 
This equation is the discrete analogous of the continuum model, see \cite{Strauss} for a review of the well-posedness of the continuum model. Equation (\ref{phi4}) can be seen as a discretisation of such a continuum model; it is therefore well-posed.
The power of the nonlinearity is important to get a well-posed problem. This can be seen by omitting the Laplacian and looking at the  differential equation $\ddot{\mathbf{u}}=- \mathbf{u}^{\alpha}$ whose solutions are bounded for $\alpha$ odd.

Since the graph Laplacian $\mathbf{\Delta}$ is a real symmetric negative-semi definite matrix, it is natural, following \cite{cks13}, to expand $\mathbf{u}$ using a basis of the eigenvectors $\mathbf{v}^{j}$ of $\mathbf{\Delta}$, such that
\begin{equation}
\label{diagonalisation1}
\mathbf{\Delta} \mathbf{v}^{j}=-\omega _{j}^{2} \mathbf{v}^{j}.
\end{equation}
The vectors $\mathbf{v}^{j}$ can be chosen to be orthonormal with respect to the scalar product in $\mathbb{R}^{N}$, i.e.  
$\prodscal{\mathbf{v}^{i}}{\mathbf{v}^{j}} = \delta_{i,j}$ where $\delta_{i,j}$ is the Kronecker symbol, so the matrix $\mathbf{P}$ formed by the columns $\mathbf{v}^{j}$ is orthogonal. The relation (\ref{diagonalisation1}) can then be written
\begin{equation*}
\mathbf{\Delta} \mathbf{P}=\mathbf{P} \mathbf{D} ,
\end{equation*}
where $\mathbf{D}$ is the diagonal matrix of diagonal $-\omega_1^2=0 > -\omega_2^2 \geq \cdots \geq -\omega_N^2$. 
The first eigenfrequency $\omega _{1}=0$ (the graph is connected) 
corresponds to the 
Goldstone mode \cite{Scott2} whose components are equal on a network $\mathbf{v}^1 = {1\over \sqrt{N}}(1,1, \dots , 1 )^T$.
We introduce the vector $\mathbf{a}= (a_1,a_2,..,a_N)^T$ such that
\begin{equation}
\label{modes_propres}
\mathbf{u}=\mathbf{P}\mathbf{a}=\sum_{k=1}^{N} a_k \mathbf{v}^k .
\end{equation}
In terms of the coordinates $a_j$, substituting (\ref{modes_propres}) into (\ref{phi4}) and projecting on each mode 
$\mathbf{v}^{j}$, we get 
the system of $N$-coupled ordinary differential equations 
\begin{equation*}
\ddot{a}_{j}=-\omega_{j}^{2} a_{j} - \sum_{m=1}^{N} u_m^3 v_m^j , ~~ j \in \{1,...,N \} ,
\end{equation*}
where we have used the orthogonality of the eigenvectors of $\mathbf{\Delta}$, $\mathbf{P}^{-1}=\mathbf{P}^{T}$. The term $u_{m}^{3}$ can be 
written as 
$u_m^3=\sum_{k,l,p=1}^{N}a_{k}a_{l}a_{p}v_{m}^{k}v_{m}^{l}v_{m}^{p}$.  
We get then a set of $n$ second order inhomogeneous coupled differential equations:
\begin{equation}
\label{eq_amplitude}
\ddot{a}_{j}+\omega_{j}^{2} a_{j}=-\sum_{k,l,p=1}^{N} \Gamma_{j k l p} a_{k}a_{l}a_{p},
\end{equation}
where 
\be \label{Gamma}
\Gamma_{j k l p}= \sum_{m=1}^{N} {v}^{j}_m~ {v}^{k}_m~{v}^{l}_m~{v}^{p}_m .
\ee

Notice that the graph geometry comes through the coefficients 
$\Gamma_{j k l p}$. 
For a general graph, the spectrum needs to be computed numerically and
these coefficients as well.
For cycles and chains however, the eigenvalues and the eigenvectors 
of the Laplacian have an explicit formula (see Appendix \ref{Spectrum}) so that $\Gamma_{j k l p}$ can be computed explicitely.
Then, we obtain the amplitude equations coupling the normal modes.

In our previous study \cite{cks13}, we noted the importance of
nodes for which the component of the eigenvector is zero. 
We introduced "a soft node" as : 
a node $s$ of a graph is a soft node for an eigenvalue 
$-\omega_j^2$ of the graph Laplacian if there exists an eigenvector 
$\mathbf{v}^j$ for this eigenvalue such that ${v}^j_s = 0$. On such
soft nodes, any forcing or damping of the system is null for the
corresponding normal mode \cite{cks13}.

\section{Existence of periodic orbits}

In \cite{aoki}, Aoki studied nonlinear periodic orbits for
chain or cycle graphs, i.e. one dimensional lattices
with free or periodic boundary conditions. He
identified a criterion allowing to extend some linear normal modes 
into nonlinear periodic orbits for the $\phi^4$ model (cubic nonlinearity). In 
this section, we generalize Aoki's criterion to any graph.

Let us find the conditions for the existence of a
nonlinear periodic solution of (\ref{phi4}) of the form
\begin{equation}
\label{ansatz}
\mathbf{u}(t)=a_j(t) \mathbf{v}^j ,
\end{equation}
the equations of motion (\ref{phi4}) reduce to
\begin{equation}\label{at}
\ddot{a}_j v_m^j =-\omega_j^2 a_j v_m^j  - a_j^3 (v_m^j)^3 .
\end{equation}
These equations are satisfied for the nodes $m$ such that $v_m^j=0$ (the soft nodes).\\
For $v_m^j \neq 0$, we can simplify (\ref{at}) by $v_m^j$ and obtain
$$\ddot{a}_j =-\omega_j^2 a_j - a_j^3 (v_m^j)^2 .$$
These equations should be independant of $m$ and this
imposes 
\begin{equation}
\label{cond}
(v_m^j)^2 =C ,
\end{equation} 
where $C$ is a constant. Remembering that $\norm{\mathbf{v}^j }=1 $
we get
$$C = {1\over N-S},$$
where $S$ is the number of soft nodes.

To summarize, for a general network, we identified 
nonlinear periodic orbits $\mathbf{u}^j(t)$ associated to
a linear eigenvector $\mathbf{v}^j$ of the Laplacian; they are 
\begin{equation}
\left\{
\begin{array}{l c r}
\label{nlp} 
\mathbf{u}^j(t) = a_j(t) \mathbf{v}^j ,\\
\\
{1 \over \sqrt{C}} v_m^j \in \{0, 1, -1\}, ~~~\forall m \in \left\{1,\dots,N \right\},~~C = {1\over N-S} ,\\
\\
\ddot{a}_j =-\omega_j^2 a_j - C a_j^3 . 
\end{array}
\right.
\end{equation}
There are three kinds of nonlinear periodic orbits 
monovalent, bivalent and trivalent depending whether they contain
one, two or three different values.
The only monovalent orbit is the Goldstone mode.
The bivalent orbits contain $+1,-1$ up to a normalization constant.
Finally, the trivalent orbits are composed of $0,+1,-1$; they posess soft nodes.

Several remarks can be made.
The first one is that the criterion can be generalized to any odd
power of the nonlinearity. We can use the condition (\ref{cond}) to 
systematically find periodic orbits for a general nonlinear wave equation with 
a polynomial nonlinearity with odd powers $\mathcal{N}(\mathbf{u})=-\mathbf{u}^3-\mathbf{u}^5 \dots$. \\
We have tried to obtain nonlinear quasi-periodic orbits of the form
$$\mathbf{u}(t) = a_j(t) \mathbf{v}^j + a_k(t) \mathbf{v}^k . $$
Preliminary results are shown in the Appendix \ref{2mode}.

\section{Examples of nonlinear periodic orbits }
\label{exp}

There are a number of examples which can be easily identified. 
First, consider the modes without soft nodes. 
\begin{itemize}

\item The monovalent mode, also named zero linear frequency mode 
or Goldstone mode
\begin{equation*}
v_m^1={1 \over \sqrt{N}}, ~~~\forall m \in \left\{ 1,\dots, N\right\} ,
\end{equation*}
exists for any graph.

\item The bivalent mode 
\begin{equation*}
v_m^j=\pm {1 \over \sqrt{N}} ,~~~\forall m\in \left\{ 1,\dots, N\right\} ,
\end{equation*}
exists for chains with $N$ even and for $j={N\over 2} +1$,
\begin{equation*}
v_m^{{N\over 2} +1} = {1\over \sqrt{N}}
\left\{
\begin{array}{l c r}
(-1)^{m \over 2}, \quad \text{if } m \text{ even},\\
\\
(-1)^{m-1 \over 2}, \quad \text{if } m \text{ odd},\\
\end{array}
\right.
\end{equation*}
corresponds to the frequency $\omega_{{N\over 2} +1}=\sqrt{2}$. 
For example, for a chain of $N=4$ nodes, we have the following
bivalent mode

\begin{figure}[H]
 \centerline{\resizebox{4 cm}{0.6 cm}{\includegraphics{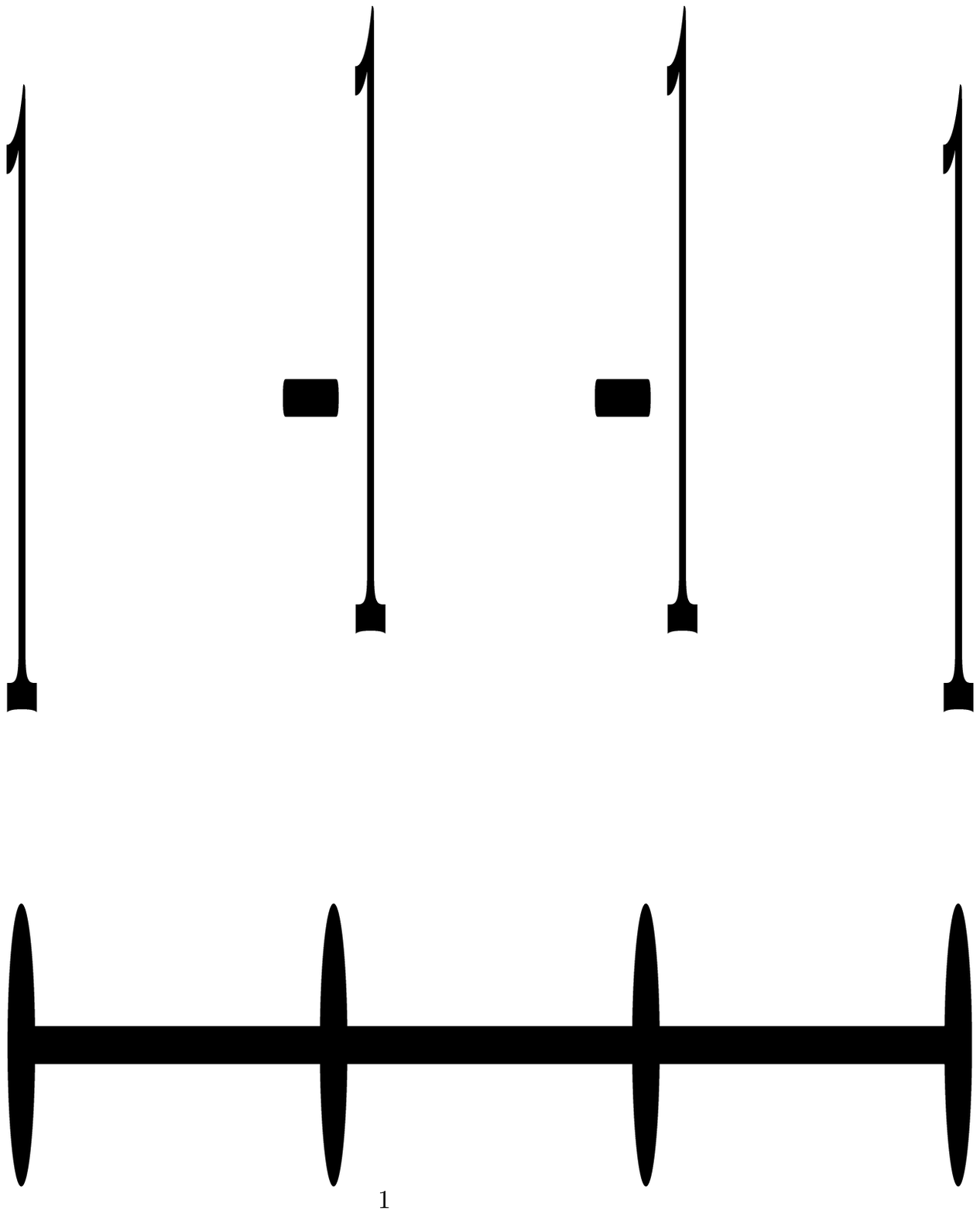}}}
\end{figure}

\item For cycles with $N$ even, the bivalent mode $\mathbf{v}^{N}$ 
alternates, it is such that $v_m^N=-v_{m+1}^N$. It corresponds 
to the frequency $\omega_{N}=2$ 
\begin{equation*}
v_m^N={1\over \sqrt{N}}(-1)^{m-1},~~~\forall m \in \left\{1,\dots,N \right\}.
\end{equation*}
For a cycle of $N=4$ nodes, we have the following bivalent mode

\begin{figure}[H]
 \centerline{\resizebox{3 cm}{3 cm}{\includegraphics{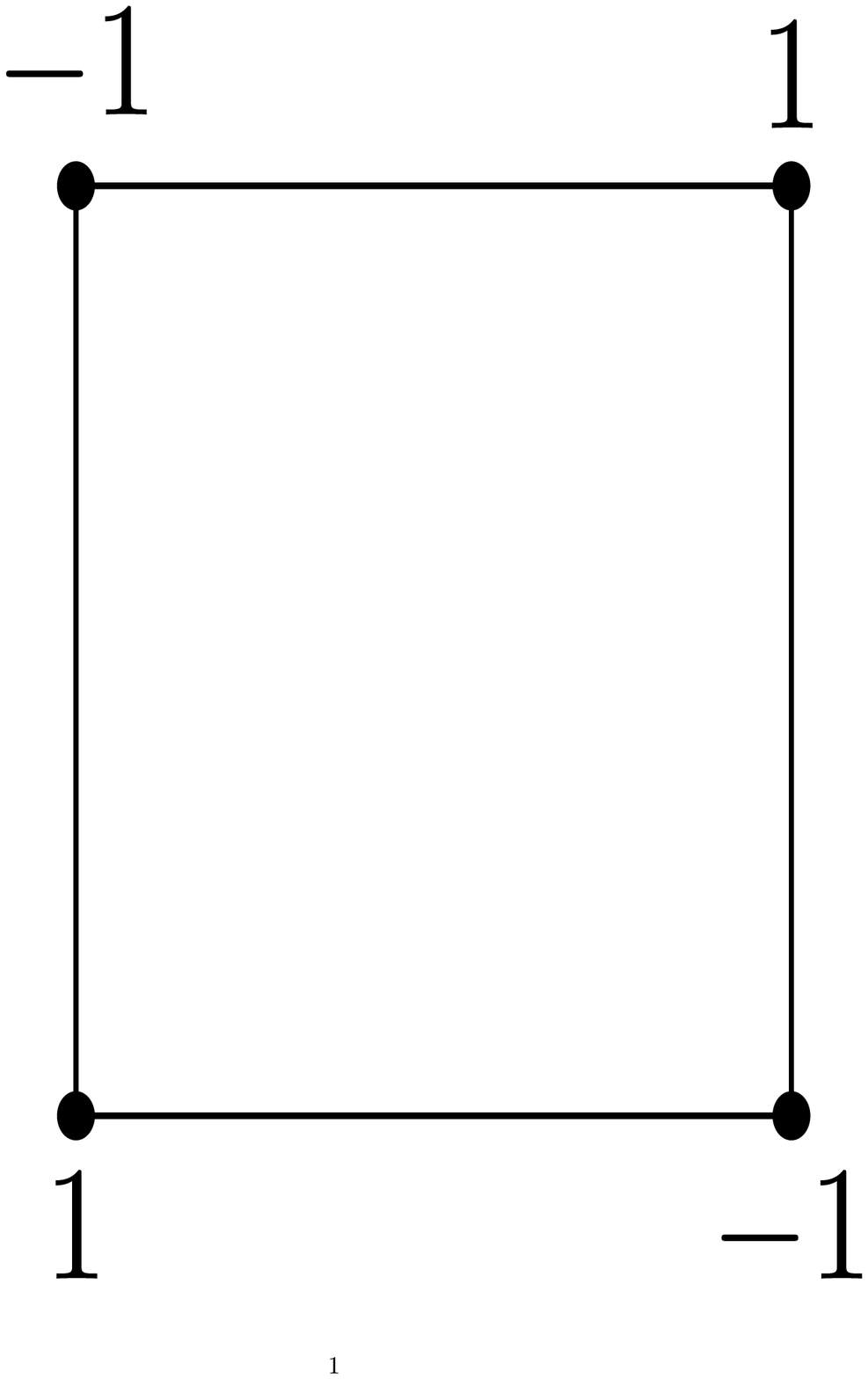}}}
\end{figure}

\end{itemize}

Other graphs that are neither a chain or a cycle can exhibit
bivalent nonlinear modes.  These are for example, using
the classification of \cite{Cvetkovic1}
\begin{itemize}
\item The Network 110 labeled as
\begin{figure}[H]
\centerline{\resizebox{5 cm}{2.5 cm}{\includegraphics{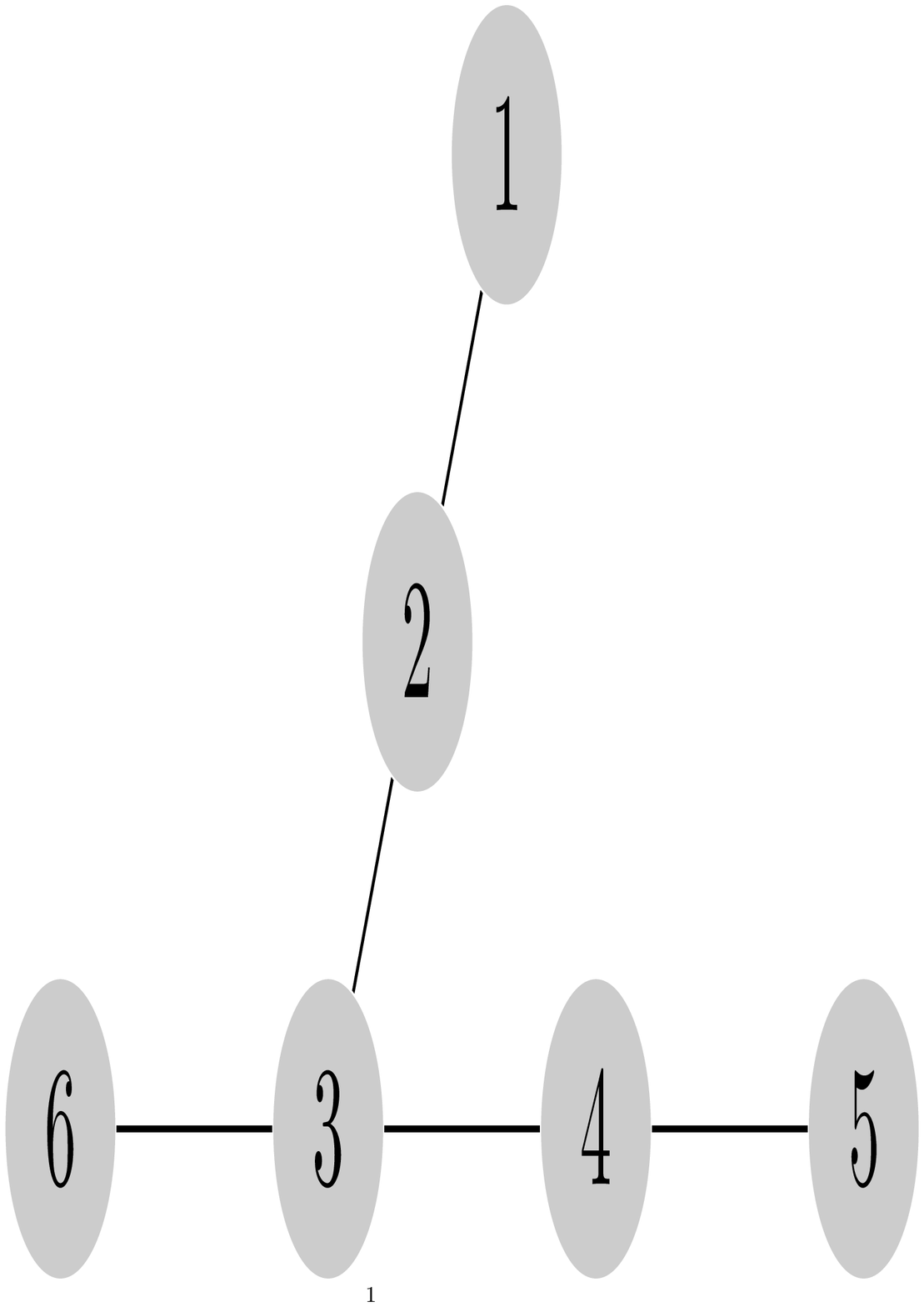}}}
\end{figure}
The nonlinear periodic orbit originates from the linear
mode, $$\omega_4^2=2,~\mathbf{v}^4={1\over \sqrt{6}}(1,-1,-1,-1,1,1)^T .$$

\item The network 105
\begin{figure}[H]
 \centerline{\resizebox{5.5 cm}{3.5 cm}{\includegraphics{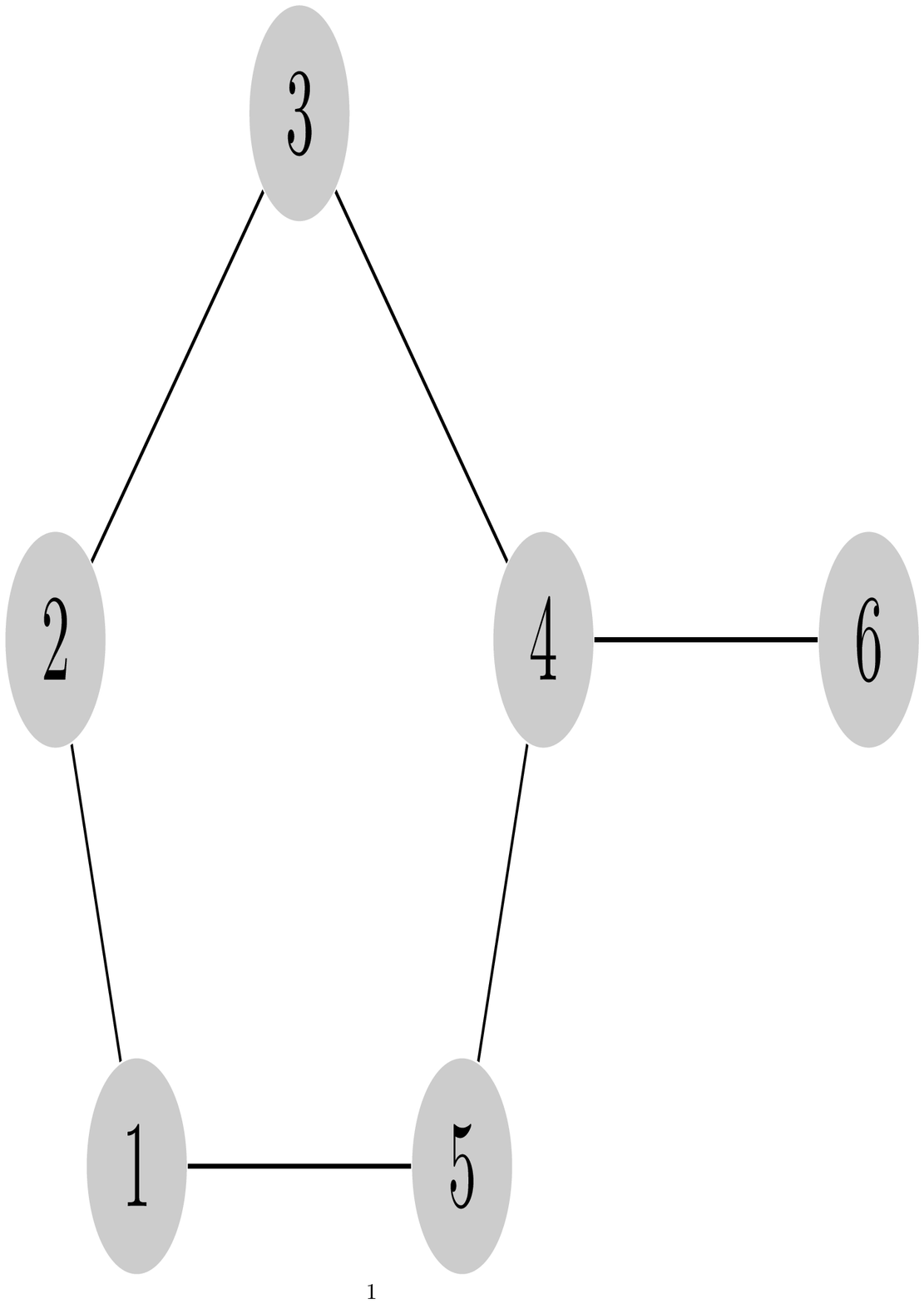}}}
\end{figure}
The nonlinear periodic orbit originates from the linear
mode, $$\omega_4^2=2,~\mathbf{v}^4={1\over \sqrt{6}}(-1,-1,1,1,1,-1)^T .$$
\end{itemize}

\noindent Nonlinear modes containing soft nodes, or trivalent modes can be found in
the following graphs:
\begin{itemize}

\item For chains with $N$ multiple of $3$, the 
mode $\mathbf{v}^{{N\over 3} +1}$ corresponds to the frequency
$\omega_{N}=1$
\begin{equation*} 
\sqrt{{2 N \over 3}} v_m^{{N\over 3} +1}\in \{0, 1, -1\},~~~\forall m \in \left\{1,\dots,N \right\}.
\end{equation*}
Notice that $v_m^{{N\over 3} +1}=0$ for $m=3k+2$ and $ k\in \{0,\dots,{N\over 3} -1 \}$.

\item For cycles where $N$ is multiple of $4$, we have a double
frequency $\omega_{N\over2} = \omega_{{N\over2}+1}=\sqrt{2}$ and two eigenvectors 
\begin{equation*}
v_m^{{N\over 2}} = \sqrt{2\over N}
\left\{
\begin{array}{l c r}
0, \quad \text{if } m \text{ even} ,\\
\\
(-1)^{m-1 \over 2}, \quad \text{if } m \text{ odd} .\\
\end{array}
\right.
\end{equation*}

\begin{equation*}
v_m^{{N\over 2}+1} = \sqrt{2\over N}
\left\{
\begin{array}{l c r}
(-1)^{{m \over 2}+1}, \quad \text{if } m \text{ even} ,\\
\\
0, \quad \text{if } m \text{ odd} .\\
\end{array}
\right.
\end{equation*}

\item For cycles with $N$ multiple of $3$, the 
mode $\mathbf{v}^{{2N\over 3} +1}$ corresponds to the double frequency
$\omega_{{2N\over 3} }=\omega_{{2N\over 3} +1}=\sqrt{3}$
\begin{equation*}
\sqrt{{2 N \over 3}} v_m^{{2N\over 3} +1}\in \{0, 1, -1\},~~~\forall m \in \left\{1,\dots,N \right\}.
\end{equation*}
Notice that $v_m^{{2N\over 3} +1}=0$ for $m=3k+1$ and $ k\in \{0,\dots,{N\over 3} -1 \}$.

\item For cycles with $N$ multiple of $6$, the 
mode $\mathbf{v}^{{N\over 3} +1}$ corresponds to the double frequency
$\omega_{{N\over 3} }=\omega_{{N\over 3} +1}=1$
\begin{equation*}
\sqrt{{2 N \over 3}} v_m^{{N\over 3} +1}\in \{0, 1, -1\},~~~\forall m \in \left\{1,\dots,N \right\}.
\end{equation*}
Notice that $v_m^{{N\over 3} +1}=0$ for $m=3k+1$ and $ k\in \{0,\dots,{N\over 3} -1 \}$.

\end{itemize}

Other networks containing soft nodes have eigenvectors extending
into nonlinear periodic orbits. These are for example
\begin{itemize}
\item the network 20 
\begin{figure}[H]
 \centerline{\resizebox{4 cm}{2.8 cm}{\includegraphics{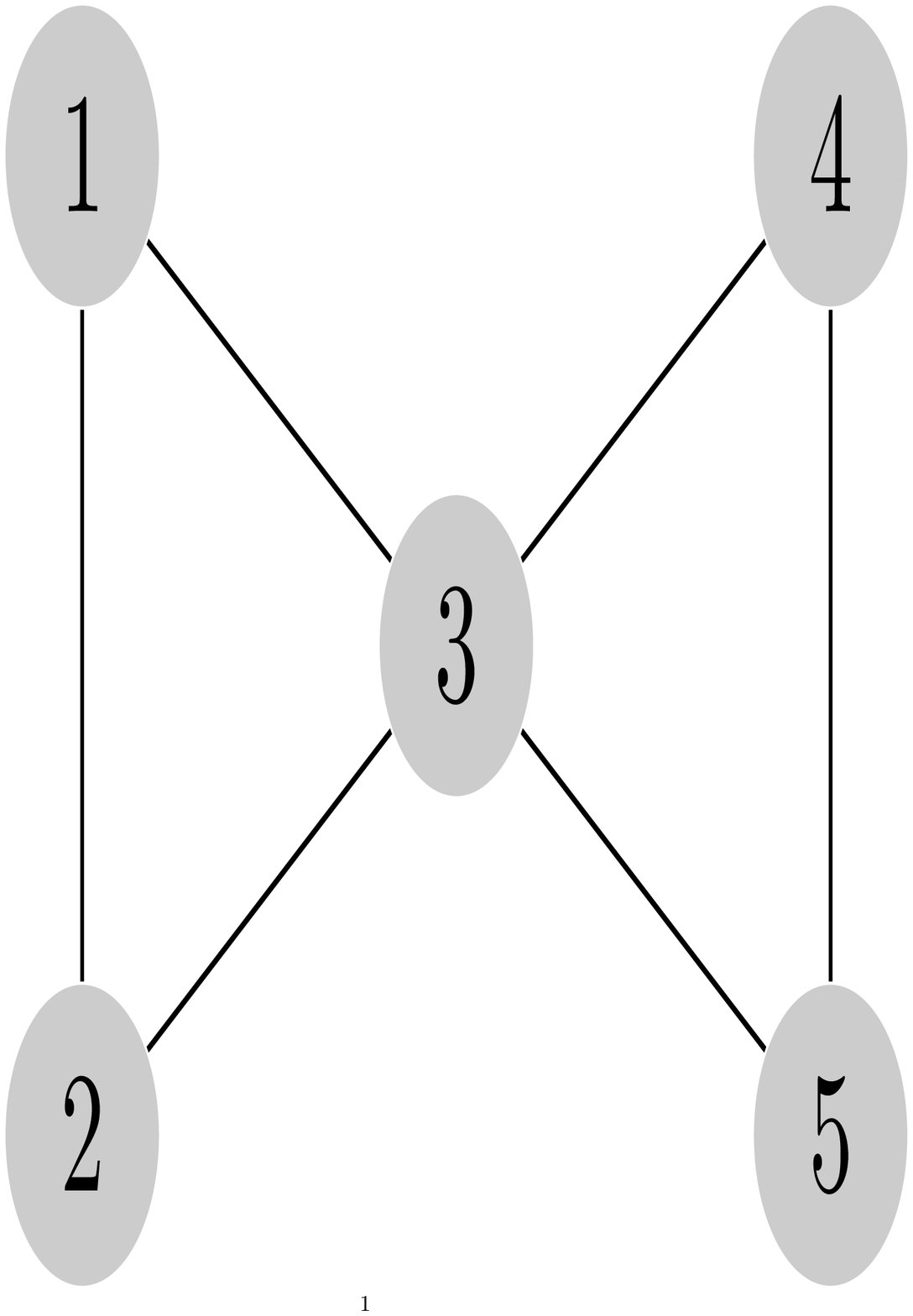}}}
\end{figure}
We have the following important parts of the spectrum
\begin{equation*}
\begin{array}{l c r}
\omega_2^2=1,~\mathbf{v}^2={1\over 2}(1,1,0,-1,-1)^T ,\\
\\
\omega_3^2=3,~\mathbf{v}^3={1\over \sqrt{2}}(-1,1,0,0,0)^T ,\\
\\
\omega_4^2=3,~\mathbf{v}^4={1\over \sqrt{2}}(0,0,0,-1,1)^T.
\end{array}
\end{equation*}

\item the network 102
\begin{figure}[H]
 \centerline{\resizebox{4 cm}{3.8 cm}{\includegraphics{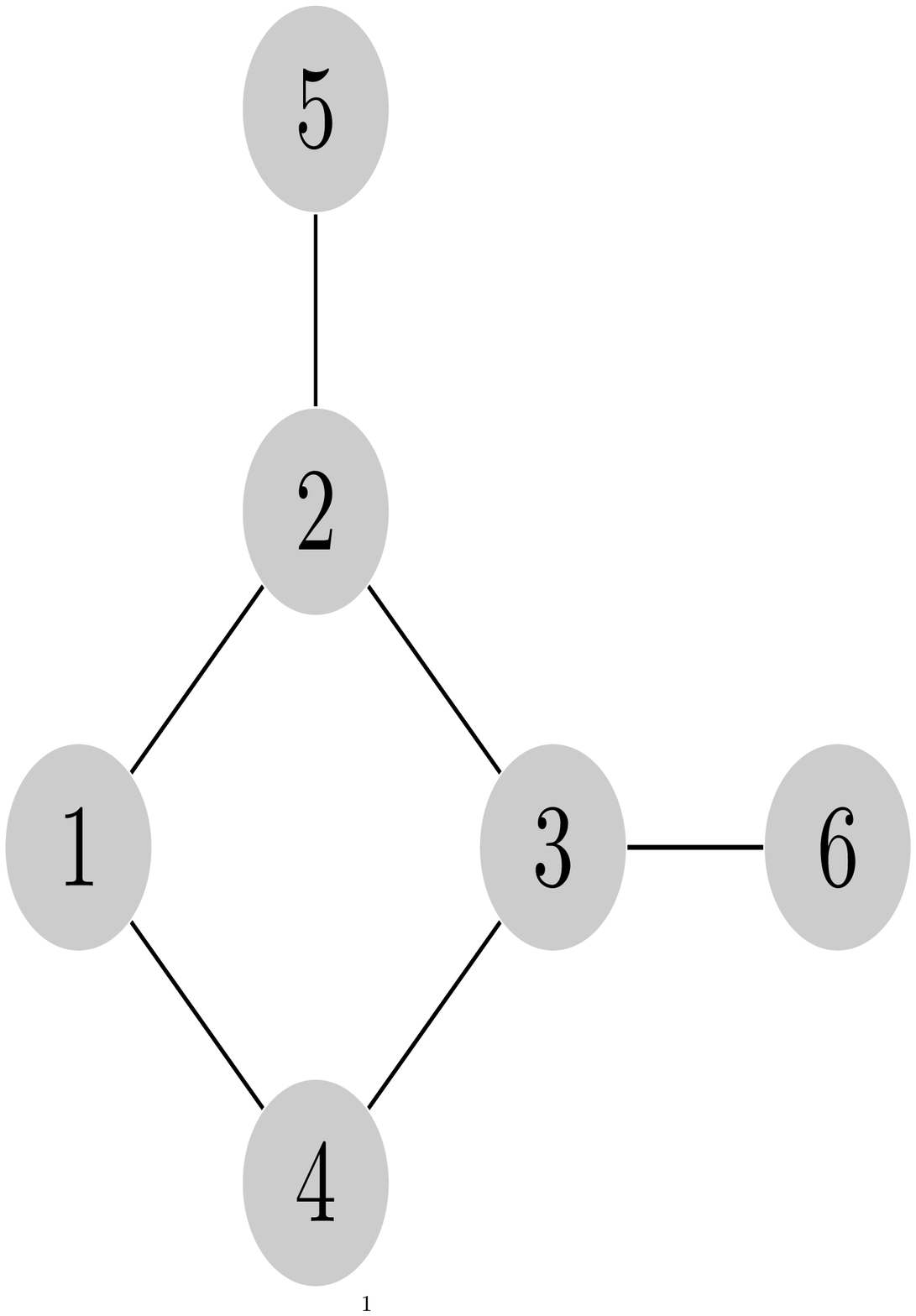}}}
\end{figure}
The nonlinear periodic orbit originates from the linear
mode, $$\omega_3^2=1,~\mathbf{v}^3={1\over 2}(1,0,0,1,-1,-1)^T .$$
\end{itemize}
See \cite{ck16} for a classification of graphs containing soft nodes.

\section{Linearization around the periodic orbits }

We now analyze the stability of the nonlinear periodic orbits
that we found by perturbation. 
This analysis reveals two main classes of orbits
depending whether they contain soft nodes or not.

To analyse the stability of (\ref{nlp}), we perturb a nonlinear 
mode $\mathbf{w}= a_j(t)\mathbf{v}^j$ satisfying (\ref{nlp}) and write
$$\mathbf{u}= \mathbf{w} +\mathbf{y},$$ where 
$\norm{\mathbf{y}}\ll \norm{\mathbf{w}}$. Plugging the above expression
into (\ref{phi4}), we get for each coordinate $i$
\begin{equation}
\label{perturb}
\ddot {y}_i 
= \sum_{k=1}^{N} \Delta_{ik} y_k -3~w_i^2 y_i -3~w_i y_i^2 - y_i^3 ,
\end{equation}
where we have used the fact that $\mathbf{w}$  is a solution of
(\ref{phi4}).

Two situations occur here, depending if the eigenvector $\mathbf{v}^j$
contains zero components (soft nodes) or not. 
If there are no soft nodes $w_i \neq 0,~~\forall i\in \left\{1,\dots,N \right\}$, like for the Goldstone mode or
the bivalent mode, equation (\ref{perturb}) can be
linearized to
\begin{equation}
\label{Hill}
\ddot{\mathbf{y}}=\mathbf{\Delta} \mathbf{y}-
{3 \over N } a_j^2(t) \mathbf{y} ~~.
\end{equation}
Expanding $\mathbf{y}$ on the normal modes, 
$\mathbf{y}=\sum_{k=1}^{N}z_k(t)\mathbf{v}^k$ we decouple (\ref{Hill}) 
and obtain $N$
one dimensional Hill-like equations for each amplitude $z_k$
\begin{equation}
\label{Hill2}
\ddot{z}_k=-\left [\omega_k^2 + {3 \over N } a_j^2(t) \right ] z_k, ~~~k\in \left\{1,\dots, N\right\} .
\end{equation}
Again we generalize the result, obtained by Aoki \cite{aoki} for chains 
and cycles, to a general graph.

In the case where there are soft nodes; 
we can also write the linearized equations.
First, let us assume for simplicity that there is only 
one zero component $m$ of $\mathbf{v}^j$, then $w_m=0$ 
so that we need to keep the cubic term $y_m^3$ in (\ref{perturb}) for $i=m$ and
we can linearize (\ref{perturb}) for all $i\neq m$. The evolution of $\mathbf{y}$ is given by 
\begin{eqnarray*}
\ddot{y}_i&=&\sum_{p=1}^{N} \Delta_{ip} y_p -{3 C } a_j^2(t) y_i, ~~i\in \left\{1,\dots,N \right\}, ~~i\neq m ,~~~\\
\ddot{y}_m&=&\sum_{p=1}^{N} \Delta_{mp} y_p- y_m^3.
\end{eqnarray*}
Expanding $\mathbf{y}$ on the normal modes, $\mathbf{y}=\sum_{l=1}^{N}z_l(t)\mathbf{v}^l$, we get
\begin{eqnarray}
\label{eqi}
\sum_{l=1}^{N}\ddot{z}_l v_i^l&=&-\sum_{l=1}^{N} \omega_l^2 z_l v_i^l-{3 C } a_j^2(t) \sum_{l=1}^{N} z_l v_i^l, ~i\neq m ,~~~\\
\label{eqm}
\sum_{l=1}^{N}\ddot{z}_l v_m^l&=&-\sum_{l=1}^{N} \omega_l^2 z_l v_m^l- \sum_{l,p,q=1}^{N}z_l z_p z_q v_m^l v_m^p v_m^q .
\end{eqnarray}
We now multiply (\ref{eqi}) by $v_i^k$ and sum over $1\leq i \leq N$ with $i\neq m$ and multiply (\ref{eqm}) by $v_m^k$.
The two equations are 
\begin{eqnarray*}
\sum_{l=1}^{N}\ddot{z}_l \sum_{{i\neq m}}v_i^l v_i^k=
-\sum_{l=1}^{N} \left( \omega_l^2 z_l  +{3~ C } a_j^2(t)z_l  \right) 
\sum_{{i\neq m}}v_i^l v_i^k, ~~~\\
\sum_{l=1}^{N}\ddot{z}_l v_m^l v_m^k =-\sum_{l=1}^{N} \omega_l^2 z_l v_m^lv_m^k- \sum_{l,p,q=1}^{N}z_l z_p z_q v_m^l v_m^p v_m^q v_m^k.
\end{eqnarray*}
Adding the above equations and using the orthogonality condition
$\sum_{i\neq m}v_i^l v_i^k=\delta_{l,k}-v_m^l v_m^k$, we obtain
\begin{equation}
\label{nonlinear_soft}
\begin{array}{r c l}
\ddot{z}_k&=&-\left(\omega_k^2 + {3 C } a_j^2(t)\right)z_k +{3 C } a_j^2(t)\sum_{l=1}^{N} z_l v_m^l v_m^k  \\
\\
&-&\sum_{l,p,q=1}^{N}z_l z_p z_q v_m^l v_m^p v_m^q v_m^k .
\end{array}
\end{equation}
Equation (\ref{nonlinear_soft}) shows that the amplitudes $z_k$ of the perturbation $\mathbf{y}$ 
around a nonlinear periodic orbit $\mathbf{w}= a_j(t)\mathbf{v}^j$ containing a soft node $v_m^j=0$,
are coupled linearly.

Omitting the nonlinear term and keeping only the linear coupling term in (\ref{nonlinear_soft}), we obtain $N$
one dimensional coupled equations for each amplitude $z_k$
\begin{equation*}
\ddot{z}_k=-\left(\omega_k^2+ {3 C } a_j^2(t) \right)z_k +{3 C } a_j^2(t) \sum_{l=1}^{N} z_l v_m^l v_m^k.
\end{equation*}

In the general case, let us denote by 
$\mathcal{S}_j=\left\{m, ~~v_m^j=0 \right\}$ the set of the soft nodes of the trivalent mode $\mathbf{v}^j$,
then the variational system can be written 
\begin{equation}
\label{linear_soft}
\left\{
\begin{array}{r c l}
\ddot{z}_k &=& -\left(\omega_k^2+ {3 C } \left(1-\sum_{m\in \mathcal{S}_{j}}(v_m^k)^2\right) a_j^2(t) \right)z_k \\
\\
&+&{3 C } a_j^2(t)  \sum_{l\neq k} 
z_l \sum_{m\in \mathcal{S}_{j}} v_m^l v_m^k,~~ \forall k\neq j ,\\
\\
\ddot{z}_j&=&-\left(\omega_j^2+ {3 C } a_j^2(t) \right)z_j.
\end{array}
\right.
\end{equation}

The linearized equations (\ref{linear_soft}) show how the modes will couple.
If $\mathcal{S}_j \subset \mathcal{S}_k$ for a such $k \in \left\{1,\dots,N \right\}, ~~ k\neq j$,
the nonlinear mode $\mathbf{v}^j$ will not couple with the mode $\mathbf{v}^k$ 
i.e. $\mathbf{v}^k$ will not be excited when exciting the mode $\mathbf{v}^j$.
Another factor of the uncoupling is when the coupling coefficients 
$\sum_{m\in \mathcal{S}_{j}} v_m^l v_m^k=0, ~~ \forall ~ l \neq k $.

To summarize, the stability of the Goldstone and the bivalent
nonlinear periodic orbits
$\mathbf{w}$, is governed by the $N$ decoupled equations (\ref{Hill2}). 
For nonlinear periodic orbits containing soft nodes (trivalent modes), the stability is
given by the coupled system (\ref{linear_soft}). In all cases, the
orbit  will be stable if the solutions $z_k$ are bounded for all $k$.

\section{Stability of the Goldstone and the bivalent periodic orbits: Floquet analysis}

The variational system corresponding to the Goldstone mode or the bivalent mode can be decomposed into the set of independant (uncoupled) equations (\ref{Hill2})
where $a_j(t)$ is the Goldstone or the bivalent periodic elliptic function solution of
\begin{equation}
\label{eq2}
\ddot{a}_j = -\omega_j^2 a_j -\frac{1}{N} a_j^3,
\end{equation}
for proper initial conditions $a_j(0)$ and $\dot{a}_j(0)$.

In order for the Goldstone mode and the bivalent mode to be stable, the solutions of the 
differential equations (\ref{Hill2}) must be bounded $\forall k \in \left\{1,\dots, N\right\}$.
The equations (\ref{Hill2}) are uncoupled Hill-like equations and can be studied separately for each $k$. 
The evolution of $z_k$ in (\ref{Hill2}) can be obtained using Floquet multipliers \cite{Meiss}; 
the latter requires the integration of the first order variational equations 
\begin{equation}
\label{variational}
\left\{
\begin{array}{l c r}
\dot{\mathbf{M}}=\mathbf{A}_k(t)\mathbf{M} ,
\\
\mathbf{M}(0)=\mathbf{I_{2}} ,
\end{array}
\right.
\end{equation}
where $\mathbf{M}$ is a $2\times 2$ matrix with column components 
$\left(z_k(t),\dot{z}_k(t)\right)^T$, $\mathbf{I_2}$ is the 
$2\times 2$ identity matrix and 
\begin{equation*}
\mathbf{A}_k(t)=
\begin{pmatrix}
0 & 1  \\
-\left(\omega_k^2+{3 \over N} a^2_j(t)\right) & 0
\end{pmatrix},
\end{equation*}
where $a_j(t)$ is the Goldstone or the bivalent periodic orbit solution of (\ref{eq2}).
The fundamental matrix solution of (\ref{variational}) is $\mathbf{M}(t)$.
For $t=T$, the period of $a_j$, the matrix $\mathbf{M}(T)$ is called the 
monodromy matrix. The eigenvalues of $\mathbf{M}(T)$ are the Floquet multipliers 
and Floquet's theorem \cite{Meiss} states that all the solutions of (\ref{variational}) are 
bounded whenever the Floquet multipliers have magnitude smaller than one. 
To calculate the Floquet multipliers, we integrate over the period $T$ the first order variational equations (\ref{variational}) 
simultaneously with the equation of motion (\ref{eq2}).
For this, we used a fourth order Runge-Kutta routine and the Matlab infrastructure \cite{matlab}.

\subsection{Goldstone periodic orbit}
\begin{figure}[H]
 \centerline{\resizebox{7.7 cm}{5.7 cm}{\includegraphics{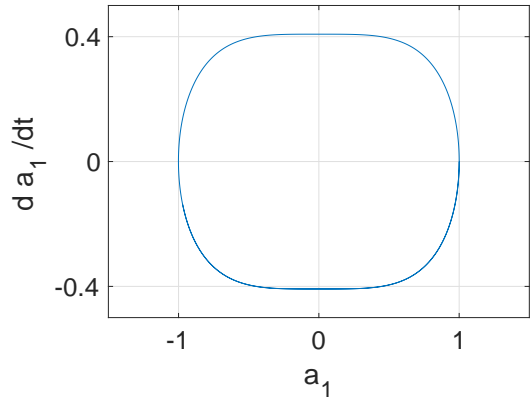}}}
 \caption{\label{orbit} Goldstone periodic orbit for $N=3$, $a_1(0)=1$ and $\dot{a}_1(0)=0$.}
\end{figure}

For a general graph  with $N$ nodes, the Goldstone periodic orbit $a_1(t)$ solution of (\ref{eq2}) for $j=1, ~ \omega_1=0$, 
can be written in terms of Jacobi elliptic functions \cite{Abramowitz}.
The solutions lie on the level curves of the energy $E={1 \over 2}(\dot{a}_1)^2 + \frac{1}{4N} a_1^4$ which is a constant of the motion.
Therefore, the phase portrait is easily obtained by plotting the level curves Fig. \ref{orbit}.
The period of oscillations is
\begin{equation*}
T = {\sqrt{N} ~\Gamma^2({1\over 4}) \over a_1(0)\sqrt{\pi}},
\end{equation*}
where $\Gamma(.)$ is the gamma function and $\Gamma({1\over 4}) \approx 3.6256$. The frequency of oscillations is
$$\omega_{\text{NL}}={2\pi \over T}={2\pi\sqrt{\pi}  \over \sqrt{N} ~\Gamma^2({1\over 4})} a_1(0).$$
We set $\gamma=\left( 4N~E\right)^{1 \over 4}$, 
then we can write the solution as 
$$a_1(t)=\gamma \mathrm{cn}(\gamma t,{1\over \sqrt{2}}),$$
where $\mathrm{cn}(t,k)$ is the cosine elliptic function 
\cite{Abramowitz} with modulus $k$ and where we have chosen $\dot{a}_1(0)=0$.

The variational equations (\ref{Hill2}) can be written for the Goldstone periodic orbit as
\begin{equation}
\label{Lame}
\ddot{z}_k=-\left(\omega_k^2+{3 \over N} \gamma^2\mathrm{cn}^2(\gamma  t,{1\over \sqrt{2}})\right)z_k , ~~~k \in \left\{1,\dots, N\right\}.
\end{equation}
Equations (\ref{Lame}) are uncoupled Lamé equations in the Jacobian form \cite{Ince} and can be studied
separately for each $k$. 
Note that the stability domain of (\ref{Lame}) 
was determined for example in \cite{Frolov};
it can be seen that there are instable bounds in the plane $(\gamma^2, \omega_k^2)$.

\subsection{The bivalent periodic orbit}
The bivalent periodic orbit $a_j(t)$ solution of (\ref{eq2}) can 
be expressed via Jacobi elliptic cosine \cite{Abramowitz}
$$a_j(t)=\gamma \mathrm{cn}(\Omega t,k^2),$$
where $\gamma=a_j(0), ~\dot{a}_j(0)=0$
and
$\Omega^2={\omega_j^2 \over 1-2 k^2},$
while the modulus $k$ of the elliptic function is determined by 
$2 k^2= {\gamma^2 \over N \omega_j^2 + \gamma^2}.$

For the bivalent periodic orbit, the variational system are uncoupled Lamé equations in the Jacobian form
\begin{equation}
\label{Lame2}
\ddot{z}_k=-\left(\omega_k^2+{3 \over N} \gamma^2\mathrm{cn}^2(\Omega  t,k^2)\right)z_k , ~~~k \in \left\{1,\dots, N\right\}.
\end{equation}

For chains, we studied systematically the Floquet multipliers 
for the Goldstone and the bivalent periodic orbits.

\subsection{Floquet analysis of Goldstone periodic orbit in chains}
\begin{figure}[H]
 \centerline{\resizebox{8.8 cm}{6.8 cm}{\includegraphics{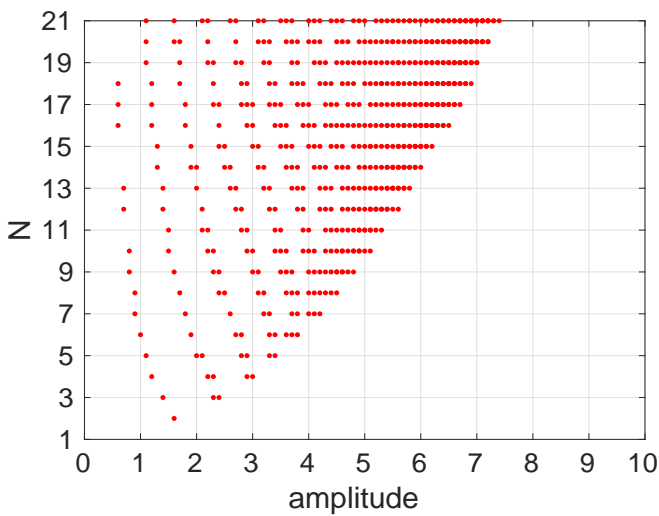}}}
       \caption{\label{stab_mode1} Instability regions of the Goldstone periodic orbit for chains with $N$ nodes for different initial amplitude $a_1(0)$.}
\end{figure}

For chains with $N$ nodes,
the instability region of the Goldstone
mode is shown in Fig. \ref{stab_mode1} as a function of
the amplitude $a_1(0)$. The points indicate instablility.
The plot shows the unstable tongues typical of the Mathieu or Lamé equations \cite{Frolov}.
For small chain sizes, there are just a few very narrow unstable tongues, for
example for $N=4$, we have three tongues. As the chain gets longer, the number of the
unstable tongues and their width increases. Note however that for large enough amplitude,
the Goldstone mode is always stable. 

\subsection{Floquet analysis of the bivalent periodic orbit in chains}
\begin{figure}[H]
 \centerline{\resizebox{8 cm}{6 cm}{\includegraphics{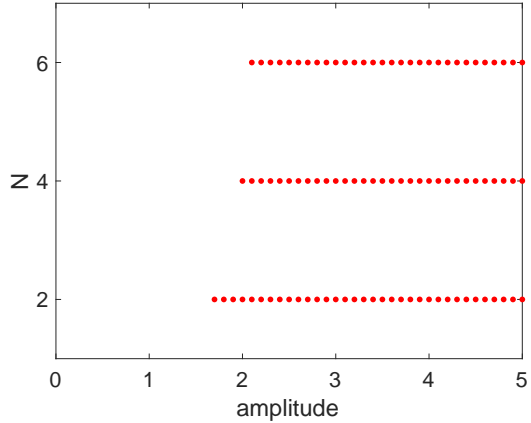}}}
        \caption{\label{bivalent} Instability regions of the bivalent 
mode $\mathbf{v}^{{N\over 2} +1}$ for chains with an even number of nodes $N$ for different amplitudes $a_{{N\over 2} +1}(0)$.}
\end{figure}

Remember that the bivalent mode only exists for chains with an even number of nodes $N$.
We calculate the instability region of the bivalent mode $\mathbf{v}^{{N\over 2} +1}$ 
and present it in Fig. \ref{bivalent} as a function of the amplitude $a_{{N\over 2} +1}(0)$. 
The points indicate the unstable solutions of (\ref{Lame2}) with $N$ even. 
For a narrow region starting from a zero amplitude, the bivalent mode is stable. Above
a critical amplitude it is unstable. Notice the difference with the Goldstone mode which is
mostly stable while the bivalent mode is mostly unstable.

To illustrate the dynamics of the Goldstone and the bivalent modes,
we solve (\ref{phi4}) for a chain of length 1.
We confirm the results of the Floquet analysis 
and show the couplings that occur in the instability regions.

\subsection{Example: Chain of length 1}
For a chain of length 1,
\begin{figure}[H]
 \centerline{\resizebox{2.2 cm}{0.6 cm}{\includegraphics{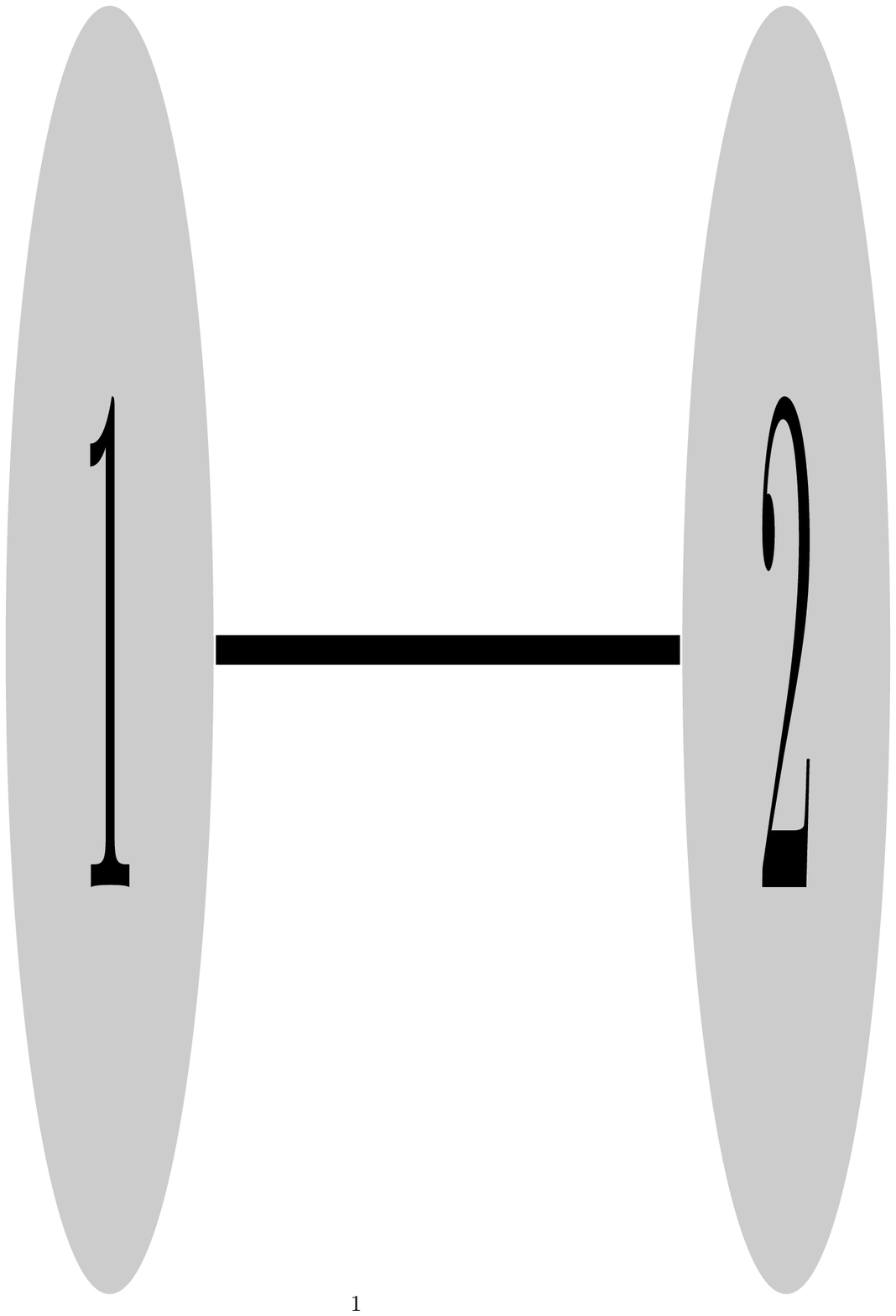}}}
\end{figure}
the spectrum is
\begin{equation*}
\begin{array}{l c r}
\omega_{1}^{2}=0,~\mathbf{v}^1={1 \over \sqrt{2}} (1,1)^T , \\
\\
\omega_{2}^{2}=2, ~\mathbf{v}^2={1 \over \sqrt{2}} (1,-1)^T .
\end{array}
\end{equation*}

The amplitude equations (\ref{eq_amplitude}) are:
\begin{equation}
\label{eqs_ch2}
\left\{
\begin{array}{r c l}
\ddot{a}_1 &=& \frac{-1}{2} a_1^3 - {3\over 2}a_1 a_2^2 , \\
\\
\ddot{a}_2 &=&  -2a_2- \frac{1}{2} a_2^3- {3\over 2}a_1^2 a_2.
\end{array}
\right.
\end{equation}

\begin{figure}[H]
\centerline{\resizebox{5.5 cm}{7.5 cm}{\includegraphics{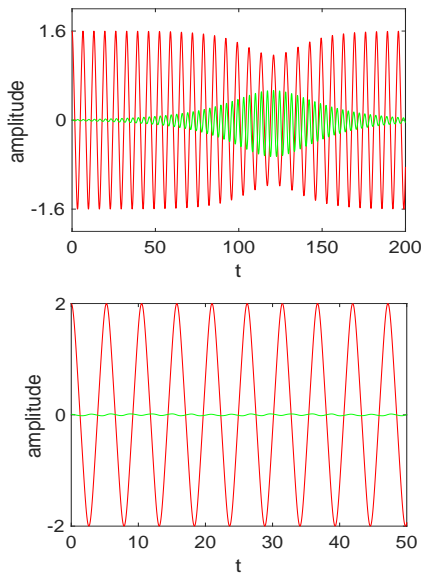}}}
  \caption{\label{m1ch2} 
Time evolution of the mode amplitudes $a_1$ (red online) and $a_2$ (green online)
when exciting Goldstone mode $\mathbf{v}^1$ for $a_1(0)=1.6$ (top) and $a_1(0)=2$ (bottom) with $a_2(0)=10^{-2}$.}
\end{figure}

First we consider the evolution of the Goldstone nonlinear periodic orbit.
We solve numerically (\ref{phi4}) for an initial condition $\mathbf{u}(0)=a_1(0)\mathbf{v}^1$ with $\dot{\mathbf{u}}(0)=0$.
The top panel of Fig. \ref{m1ch2} shows the amplitudes $a_1(t),a_2(t)$ for $a_1(0)=1.6$.
As expected from the Floquet analysis Fig. \ref{stab_mode1} the orbit is unstable 
and gives rise to a coupling with the mode $\mathbf{v}^2$.
On the other hand, for $a_2(0)=2$ the amplitudes shown in the bottom panel of Fig. \ref{m1ch2} 
do no show any coupling.
The Goldstone mode is stable as shown in Fig. \ref{stab_mode1}.

\begin{figure}[H]
\centerline{\resizebox{5.5 cm}{7.5 cm}{\includegraphics{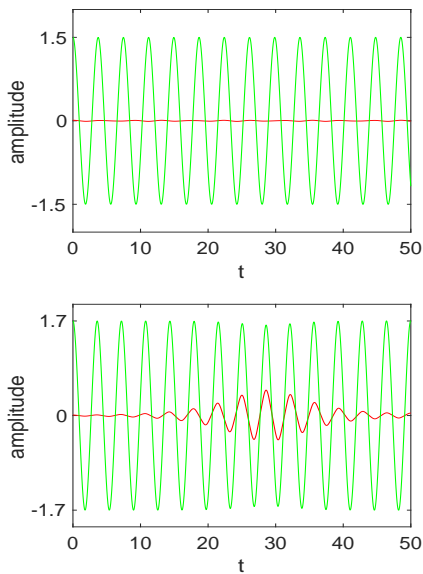}}}
    \caption{\label{m2ch2} 
Time evolution of the mode amplitudes $a_1$ (red online) and $a_2$ (green online)
when exciting the bivalent mode $\mathbf{v}^2$ for $a_2(0)=1.5$ (top) and $a_2(0)=1.7$ (bottom) with $a_1(0)=10^{-2}$.}
\end{figure}

We then consider the evolution of the bivalent mode $\mathbf{v}^2$.
For that, we solve numerically (\ref{phi4}) for an initial condition $\mathbf{u}(0)=a_2(0)\mathbf{v}^2$ 
and $\dot{a}_2(0)=0$.
For $a_2(0)=1.5$, the amplitudes shown in the top panel of Fig. \ref{m2ch2} 
do no show any coupling.
As expected from the Floquet analysis in Fig. \ref{bivalent},
the bivalent mode is stable for $a_2(0) < 1.7$ and unstable for $a_2(0)\geq 1.7$.
For $a_2(0) = 1.7$, we observe coupling to the Goldstone mode as shown in the bottom panel of Fig. \ref{m2ch2}.

\section{Nonlinear modes containing soft nodes : Numerical simulations}

To illustrate the dynamics of trivalent modes, we consider three
examples. These are the single frequency mode in a chain of length 2,
the double frequency mode of cycle 3,
and the modes of the Network 20 (classification of \cite{Cvetkovic1}).
The latter are the single frequency mode and two double frequency modes.
We show the difference in the stability of a single frequency mode 
versus a double frequency mode.

\subsection{chain of length 2}
For a chain of length 2,
\begin{figure}[H]
 \centerline{\resizebox{3 cm}{0.6 cm}{\includegraphics{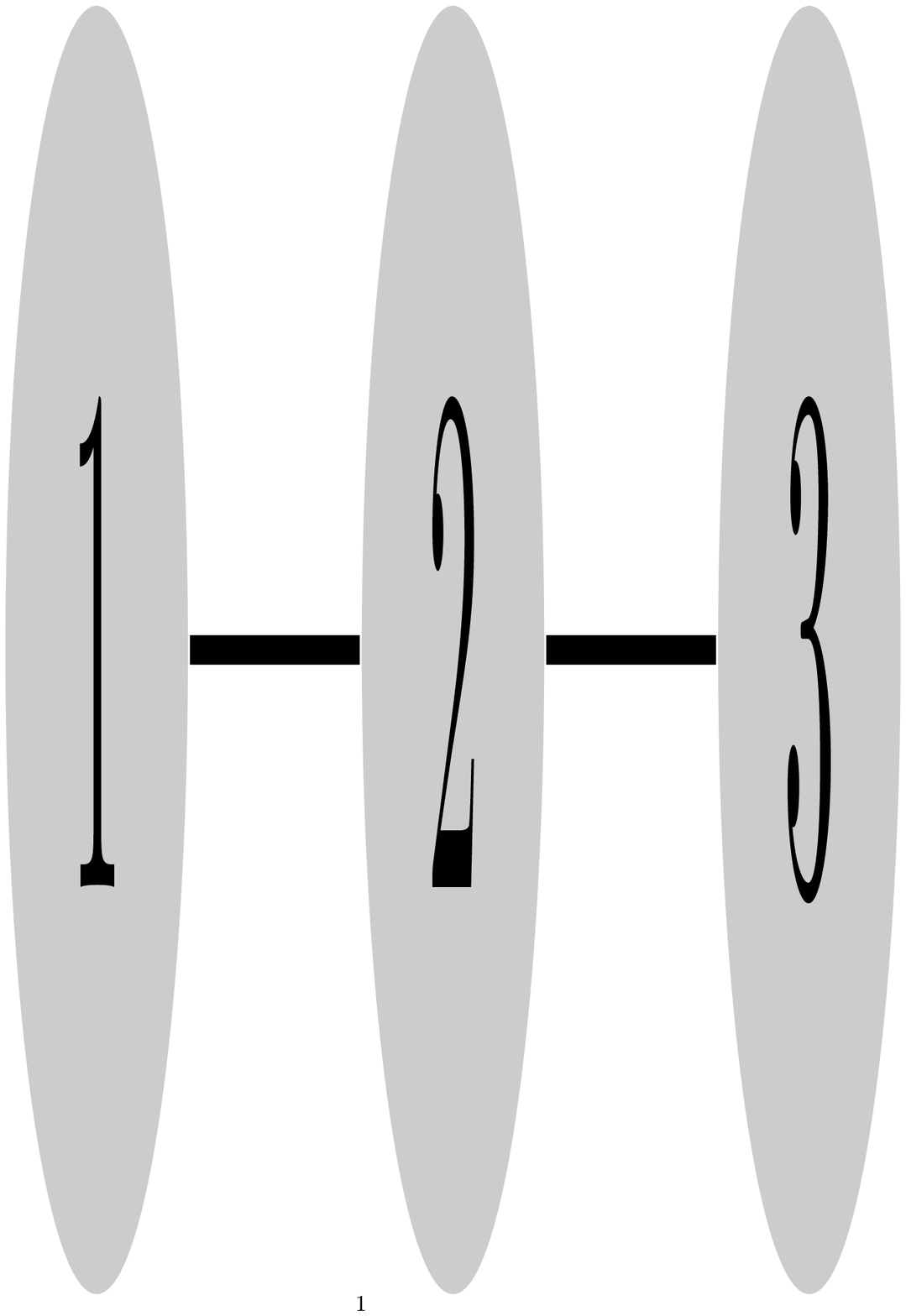}}}
\end{figure}
the spectrum is
\begin{equation*}
\begin{array}{l c r}
\omega_{1}^{2}=0,~\mathbf{v}^1={1 \over \sqrt{3}} (1,1,1)^T , \\
\\
\omega_{2}^{2}=1, ~\mathbf{v}^2={1 \over \sqrt{2}} (1,0,-1)^T , \\
\\
\omega_{3}^{2}=3, ~\mathbf{v}^3={1 \over \sqrt{6}} (1,-2,1)^T .
\end{array}
\end{equation*}

The amplitude equations (\ref{eq_amplitude}) are:
\begin{equation}
\label{eqs_ch3}
\left\{
\begin{array}{l c r}
\ddot{a}_1 = \frac{-1}{3} a_1^3 - a_1 \left( a_2^2 + a_3^2 \right) + \frac{1}{\sqrt{18}} a_3^3 
- \frac{1}{\sqrt{2}} a_2^2 a_3 ,\\
\\
\ddot{a}_2 + a_2 =   \frac{-1}{2} a_2^3-a_2 \left( a_1^2 +\frac{1}{2}  a_3^2 \right)-\sqrt{2} a_1 a_2 a_3 ,\\
\\
\ddot{a}_3 + 3 a_3 =  \frac{-1}{2} a_3^3 -a_3 \left( a_1^2 +\frac{1}{2} a_2^2 \right)-\frac{1}{\sqrt{2}}a_1 a_2^2 + \frac{1}{\sqrt{2}} a_1 a_3^2 .
\end{array}
\right.
\end{equation}

\begin{figure}[H]
\centerline{\resizebox{6 cm}{8 cm}{\includegraphics{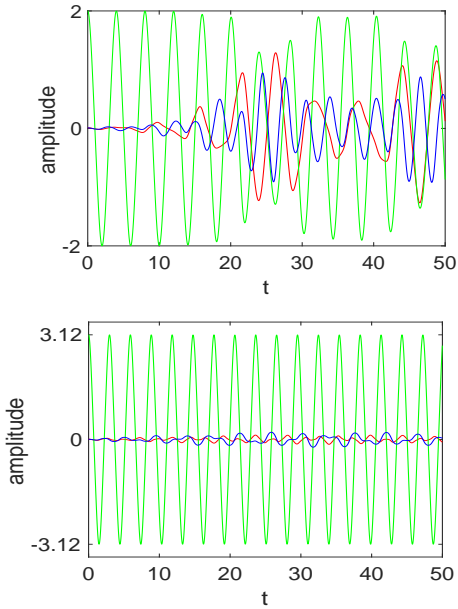}}}
       \caption{\label{m2ch3} 
Time evolution of the mode amplitudes $a_1$ (red online), $a_2$ (green online) and 
$a_3$ (blue online)
when exciting the mode $\mathbf{v}^2$ with $a_2(0)=2$ (top) and 
$a_2(0)=3.12$ (bottom). The other initial amplitudes 
are $a_1(0)=a_3(0)=10^{-2}$.}
\end{figure}

When exciting the nonlinear mode $\mathbf{v}^2$ containing a soft node, 
with $a_2(0)=2$, the modes $\mathbf{v}^1$ and $\mathbf{v}^3$ will be 
excited as shown in the top panel of Fig. \ref{m2ch3}.
This instability is explained by two factors.
First the coupling terms in the linearized equation (\ref{linear_soft}) do 
not vanish and the variational system is
\begin{equation}
\label{varch3}
{\ud^2 \over \ud t^2}
\begin{pmatrix}
z_1 \\
z_2 \\
z_3
\end{pmatrix}
=
\begin{pmatrix}
-a_2^2            & 0                                &-{1\over 2}a_2^2   \\
0                 & -\left(1 +{3\over 2}a_2^2\right) &0 \\
-{1\over 2}a_2^2  & 0                                & -\left(3+{1\over 2}a_2^2 \right)
\end{pmatrix}
\begin{pmatrix}
z_1 \\
z_2 \\
z_3
\end{pmatrix}.
\end{equation}
The second factor is the closeness of the nonlinear frequency 
$\omega_{NL} \approx 1.569$ (for $a_2(0)=2$) to the linear frequencies
of the graph. 
When the nonlinear frequency is far from the natural frequencies, 
for example when $a_2(0)\geq 3.12$ ($\omega_{NL} \geq 2.128$) the periodic 
orbit $\mathbf{v}^2$ is stable and no coupling with the other modes 
occurs as shown in the bottom panel of Fig. \ref{m2ch3}.

Notice that the matrix in the variational equations (\ref{varch3}) is the 
Jacobian matrix \cite{Meiss} of the system (\ref{eqs_ch3}) calculated 
at the periodic orbit $a_2$ solution of
$\ddot{a}_2 + a_2 =   \frac{-1}{2} a_2^3$.

\subsection{Cycle 3}
For a cycle 3,
\begin{figure}[H]
 \centerline{\resizebox{3 cm}{2.5 cm}{\includegraphics{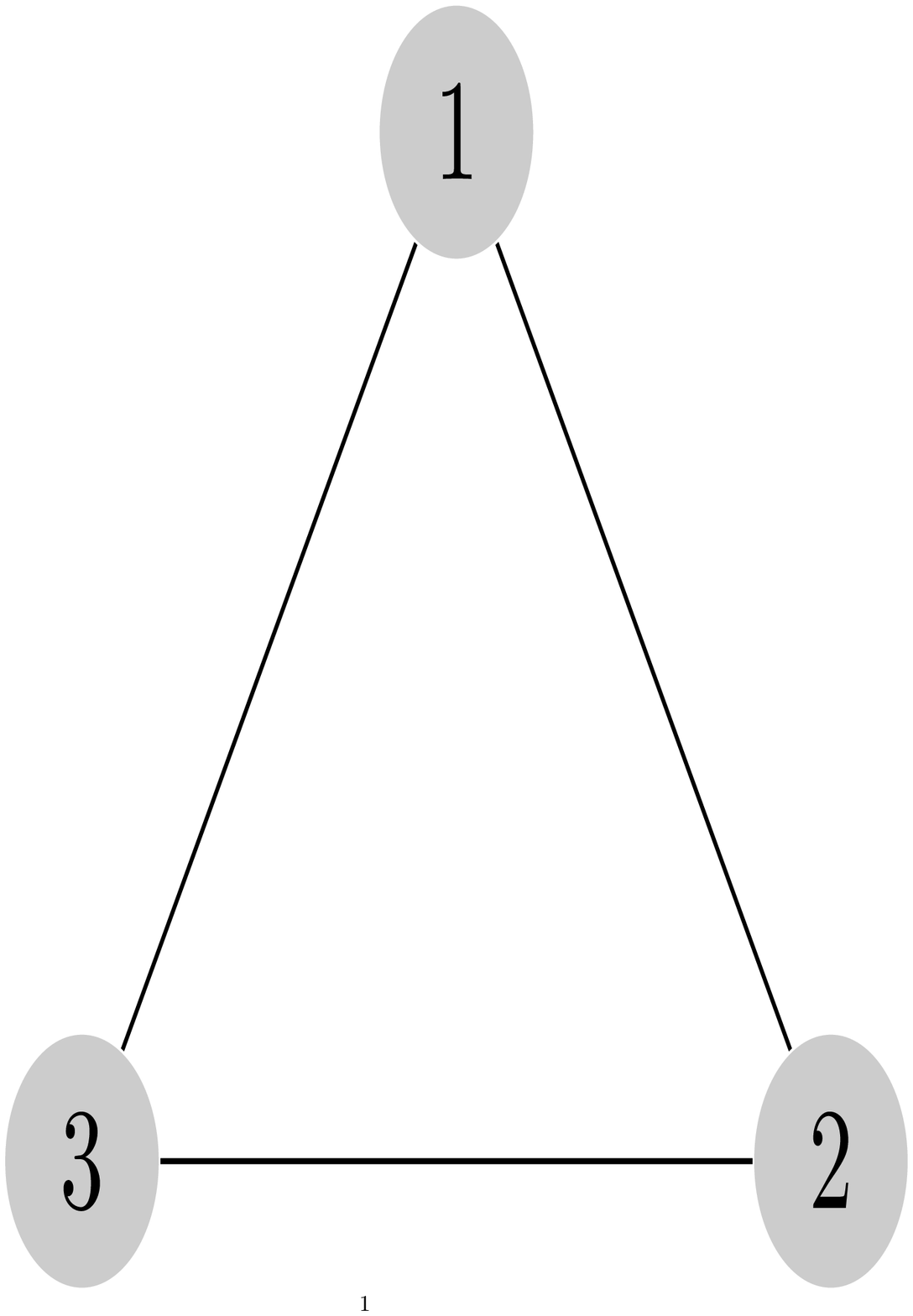}}}
\end{figure}
the spectrum is
\begin{equation*}
\begin{array}{l c r}
\omega_{1}^{2}=0,~\mathbf{v}^1={1 \over \sqrt{3}} (1,1,1)^T ,\\
\\
\omega_{2}^{2}=3, ~\mathbf{v}^2={1 \over \sqrt{6}} (2,-1,-1)^T , \\
\\
\omega_{3}^{2}=3, ~\mathbf{v}^3={1 \over \sqrt{2}} (0,1,-1)^T.
\end{array}
\end{equation*}

The amplitude equations (\ref{eq_amplitude}) are:
\begin{equation*}
\begin{array}{l c r}
\ddot{a}_1 = \frac{-1}{3} a_1^3 - a_1 a_2^2 - a_1 a_3^2 -\frac{1}{\sqrt{18}} a_2^3 + {1\over\sqrt{2}} a_2 a_3^2 ,\\
\\
\ddot{a}_2 + 3 a_2 = -a_1^2 a_2  -{1\over 2} a_2^3-{1\over 2} a_2 a_3^2  -{1\over\sqrt{2}} a_1 a_2^2 + {1\over\sqrt{2}} a_1 a_3^2 ,\\
\\
\ddot{a}_3 + 3 a_3 = -a_1^2 a_3 -{1\over 2} a_3^3 -{1\over 2} a_2^2 a_3 + \sqrt{2} a_1 a_2 a_3 .
\end{array}
\end{equation*} 

\begin{figure}[H]
\centerline{\resizebox{6 cm}{8 cm}{\includegraphics{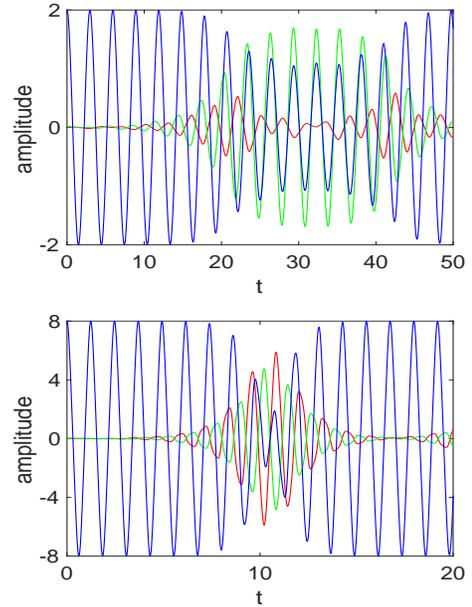}}}
       \caption{\label{m3cy3} 
Time evolution of the mode amplitudes $a_1$ (red online), $a_2$ (green online) and 
$a_3$ (blue online)
when exciting the mode $\mathbf{v}^3$ for $a_3(0)=2$ (top) and $a_3(0)=8$ 
(bottom) with $a_1(0)=a_2(0)=10^{-2}$.}
\end{figure}

The nonlinear mode $\mathbf{v}^3$ containing a soft node and 
corresponding to a double frequency,
is unstable for all initial amplitudes $a_3(0)$. It couples with the 
modes $\mathbf{v}^1$ and $\mathbf{v}^2$ as shown in Fig. \ref{m3cy3}.
There, we show the evolution of the amplitudes for 
initial conditions $a_3(0)=2$ (top) and $a_3(0)=8$ (bottom).
The coupling can be seen in the linearized equations (\ref{linear_soft})
\begin{equation*}
{\ud^2 \over \ud t^2}
\begin{pmatrix}
z_1 \\
z_2 \\
z_3
\end{pmatrix}
=
\begin{pmatrix}
-a_3^2                  &{1\over \sqrt{2}}a_3^2            & 0 \\
{1\over \sqrt{2}}a_3^2  & -\left(3 +{1\over 2}a_3^2\right) &  0 \\
0                       & 0                                & -\left(3+{3\over 2}a_3^2 \right)
\end{pmatrix}
\begin{pmatrix}
z_1 \\
z_2 \\
z_3
\end{pmatrix}.
\end{equation*}
The instability observed for large initial amplitudes seems to be
due to the degeneracy of the linear frequency as we discuss below.

\subsection{Network 20}
\begin{figure}[H]
 \centerline{\resizebox{4 cm}{2.8 cm}{\includegraphics{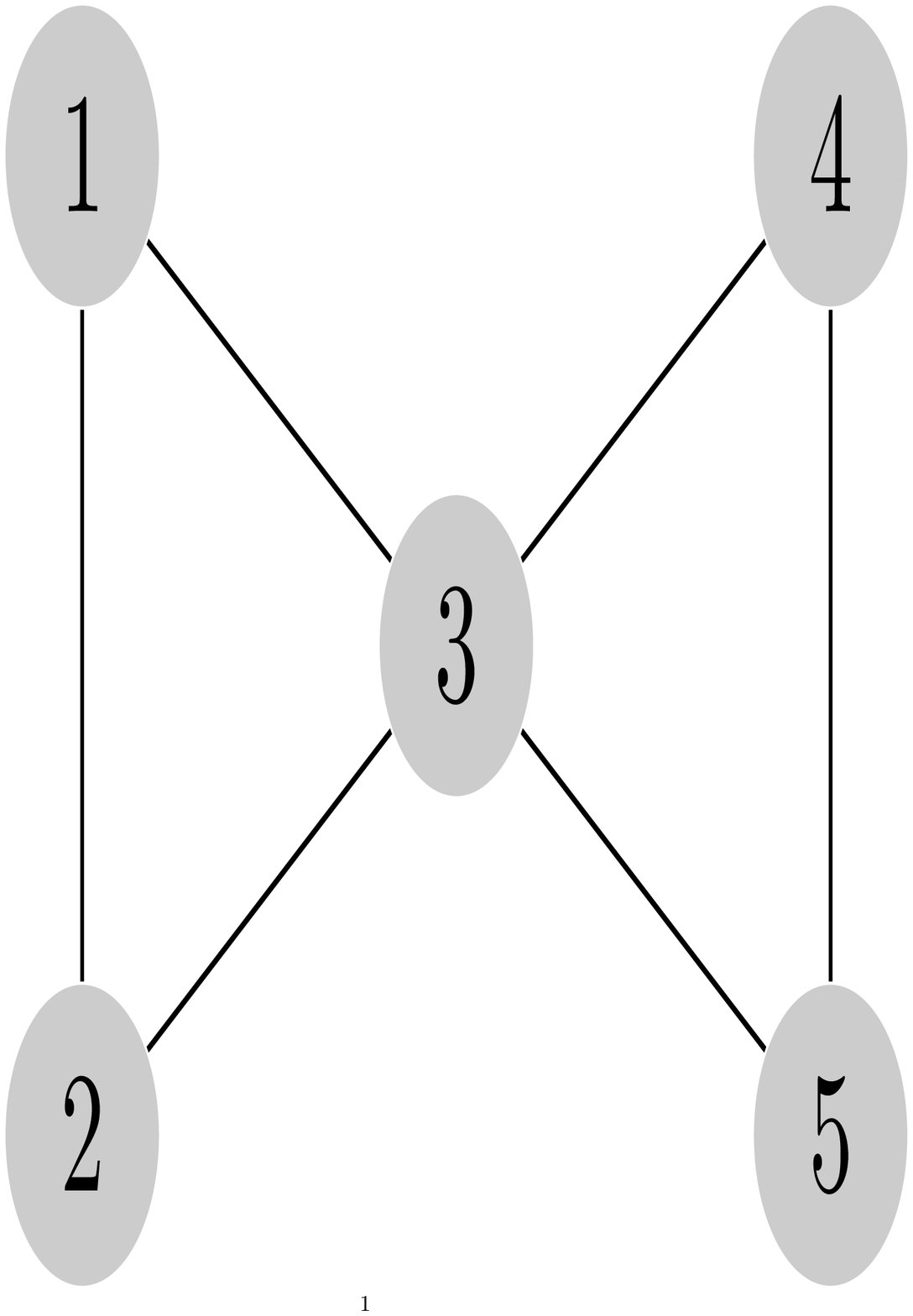}}}
\end{figure}

Network 20 contain nonlinear mode with soft node corresponding to a simple frequency,
and two nonlinear modes with soft nodes corresponding to a double frequency.
The spectrum is

\begin{equation*}
\begin{array}{l c r}
\omega_{1}=0,~\mathbf{v}^1={1 \over \sqrt{5}} (1,1,1,1,1)^T , \\
\\
\omega_2^2=1,~\mathbf{v}^2={1\over 2}(1,1,0,-1,-1)^T ,\\
\\
\omega_3^2=3,~\mathbf{v}^3={1\over \sqrt{2}}(-1,1,0,0,0)^T ,\\
\\
\omega_4^2=3,~\mathbf{v}^4={1\over \sqrt{2}}(0,0,0,-1,1)^T, \\
\\
\omega_{5}^2=5, ~ \mathbf{v}^5={1\over \sqrt{20}}(-1,-1,4,-1,-1)^T .
\end{array}
\end{equation*}

When exciting the nonlinear mode $\mathbf{v}^2$ containing a soft node 
and corresponding to a simple frequency $\omega_2=1$ for initial 
conditions $2.3 \leq a_2(0) \leq 3.37$, there is coupling with the modes $\mathbf{v}^1$ and $\mathbf{v}^5$ 
as shown in Table \ref{Net20}.
There is no coupling with $\mathbf{v}^3$ and $\mathbf{v}^4$ since the 
soft node $3$ for $\mathbf{v}^2$ is also soft for the modes
$\mathbf{v}^3$ and $\mathbf{v}^4$, so that the coupling terms 
vanish in (\ref{linear_soft}).
For $a_2(0) \geq 3.38$ ($\omega_{NL} \geq 1.755$), the nonlinear mode 
$\mathbf{v}^2$ is stable; there is no coupling with the other modes.

The nonlinear modes $\mathbf{v}^3$ and $\mathbf{v}^4$ have soft 
nodes and correspond to the double frequency 
$\omega_3=\omega_4=\sqrt{3}$. When exciting $\mathbf{v}^3$  with a 
small amplitude $a_3(0)  < 1.5$ we see no coupling with the other modes.
Starting from $a_3(0) \geq 1.5$ ($\omega_{NL} \geq 1.9587$) 
there is coupling with the modes 
$\mathbf{v}_1,~\mathbf{v}_2$ and $\mathbf{v}_5$
as shown in Table \ref{Net20} and no coupling with $\mathbf{v}^4$.
This is because the coupling terms in (\ref{linear_soft}) corresponding 
to the mode $\mathbf{v}^4$ vanish, $\sum_{m\in \mathcal{S}_{3}}  
v_m^4 v_m^k=0, ~~ \forall ~ k \neq 4 $ where $\mathcal{S}_3$ is the 
set of the soft nodes of the nonlinear mode $\mathbf{v}^3$. 
We observe similar effects when exciting $\mathbf{v}^4$ instead
of $\mathbf{v}^3$, see Table \ref{Net20}.
\begin{table}[H]
\begin{tabular}{ | l | l | l | l | l |  }
\hline
\parbox{0.12\linewidth}{Excited\\modes}&
$\omega_j$ &
\parbox{0.24\linewidth}{Nonlinear\\frequency}&
\parbox{0.24\linewidth}{amplitude for \\ instability} &
\parbox{0.2\linewidth}{Activated\\modes}  \\
\hline
$\mathbf{v}_2$ &  $1$     &   $[ 1.404 ,  1.751]$     &  $ [2.3, 3.37 ]$  &  $\mathbf{v}_2,~ \mathbf{v}_1,~\mathbf{v}_5$  \\
\hline
$\mathbf{v}_3$ &  $\sqrt{3}$     &   $ \geq 1.958$     &  $\geq 1.5$    & $\mathbf{v}_3,~ \mathbf{v}_1,~\mathbf{v}_2,~ \mathbf{v}_5$    \\
\hline
$\mathbf{v}_4$ &  $\sqrt{3}$     &   $\geq  1.958$     &   $\geq  1.5$    & $\mathbf{v}_4,~ \mathbf{v}_1,~\mathbf{v}_2,~ \mathbf{v}_5$  \\
\hline
\end{tabular}
\caption{\label{Net20} Excited modes and their associated linear frequencies, nonlinear frequencies depending on 
the initial amplitudes for instability region, and the activated modes.}
\end{table}

To summarize, we observe that a trivalent periodic orbit is stable
for large amplitudes when the eigenvalue is simple. Conversely,
when the eigenvalue is double, the periodic orbit is unstable.
In fact, the criterion of Aoki \cite{aoki} is realized only for a particular choice
of eigenvectors. Rotating the eigenspace will break the criterion
and destroy the periodic orbits. In that sense, a trivalent
periodic orbit for a multiple eigenvalue is structurally unstable.

\section{Conclusions}

The graph wave equation arises naturally from conservation laws
on a network. There, the usual continuum Laplacian is replaced by
the graph Laplacian. We consider such a wave equation with a cubic
non-linearity on a general network. 
We identified a criterion allowing to extend some linear normal modes 
of the graph Laplacian into nonlinear periodic orbits.
Three different types of periodic orbits were found, the monovalent,
bivalent and trivalent ones depending whether they contain $1$ or $-1,+1$
or $-1,0,+1$. For the monovalent and bivalent modes, the linearized
equations decouple into $N$ Hill-like equations. For chains, the 
monovalent mode is mostly stable while the bivalent is unstable.

Trivalent modes contain soft nodes and the variational equations
do not decouple. The stability is governed by a system of 
coupled resonance equations;
they indicate which modes will be excited when the orbit is unstable.
Modes that share a soft node with a trivalent orbit will not be excited.
Numerical results show that trivalent modes with a single eigenvalue are
unstable below a treshold amplitude. Conversely, trivalent modes with 
multiple eigenvalues seem always unstable.

This study can be applied to complex physical networks, 
like coupled mechanical systems.

\begin{appendices}
\section{Spectrum of cycles and chains }
\label{Spectrum}
\subsection*{Spectrum of cycles}
\begin{figure}[H]
 \centerline{\resizebox{5 cm}{4 cm}{\includegraphics{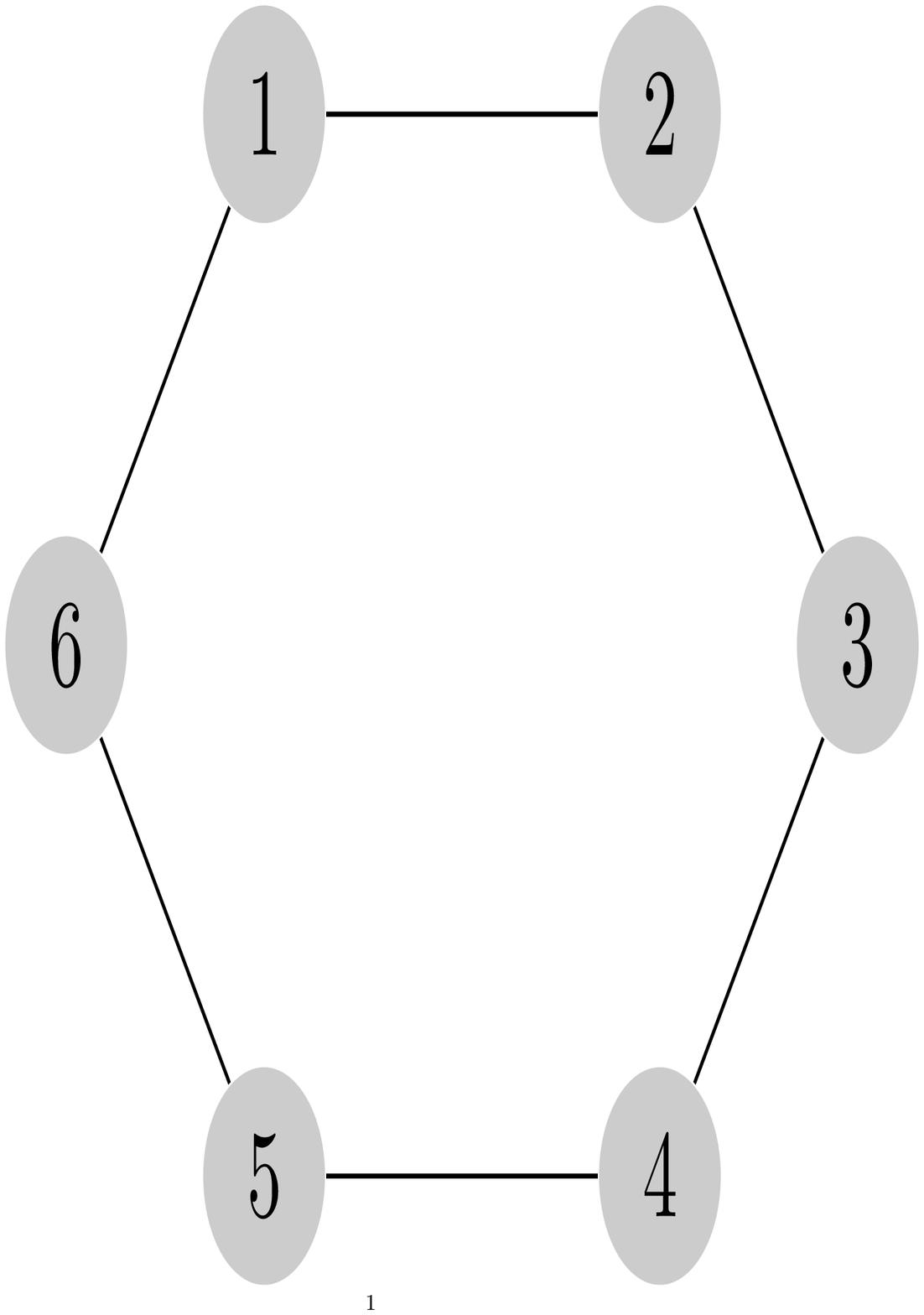}}}
 \caption{Cycle 6}
\end{figure}

For cycles, the Laplacian $\mathbf{\Delta}$ in (\ref{phi4}) is a circulant matrix \cite{Cvetkovic2} where each row vector is rotated one element to the right relative to the preceding row vector. 

\begin{equation*}
\mathbf{\Delta}=
\begin{pmatrix}
-2      &  1     &   0    & \dots  & 0       & 1    \\
1       & -2     &   1    & 0      & \dots   & 0     \\
0       & \ddots & \ddots & \ddots &  \ddots & \vdots \\
\vdots  & \ddots & \ddots & \ddots & \ddots  & 0     \\
0       & \dots  &   0    & 1      & -2      & 1      \\
1       & 0      & \dots  & 0      & 1       & -2
\end{pmatrix}
\end{equation*}

The repeated eigenvalues are 
\begin{equation*}
\begin{array}{l c r}
-\omega_{2k}^2=-\omega_{2k+1}^2=-4\sin^2\left(\frac{k\pi}{N}\right), 
\end{array}
\end{equation*}
for $k=1,\dots,\frac{N-1}{2}$ (resp. $k=1,\dots,\frac{N-2}{2}$) if $N$ is odd (resp. $N$ is even).
The first eigenvalue $-\omega_1^2=0$ is simple. When $N$ is even, the last one $-\omega_N^2=-4$ is also simple. 
The components of Goldstone eigenvector
$v_m^1=\frac{1}{\sqrt{N}}, ~ m \in \{1,...,N\}$.
We present in the following the components of the corresponding orthonormal eigenvectors
for $j\in \left\{2,\dots,N \right\}$
\begin{equation*} 
v_m^j=
\left\{
\begin{array}{l c r}
\sqrt{2\over N} \cos\left(\frac{j\pi}{N} (m-1)\right) , m\in \left\{1,\dots,N\right\}, ~j \text{ even},\\
\\
\sqrt{2\over N} \sin\left(\frac{(j-1)\pi}{N} (m-1)\right)  , m\in \left\{1,\dots,N\right\}, ~j \text{ odd} .
\end{array}
\right.
\end{equation*}

\subsection*{Spectrum of chains}
\begin{figure}[H]
 \centerline{\resizebox{4 cm}{0.6 cm}{\includegraphics{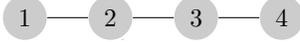}}}
 \caption{Chain of length 3}
\end{figure}

For a chain of length $N-1$ (with $N$ nodes), the Laplacian matrix is

\begin{equation*}
\mathbf{\Delta}=
\begin{pmatrix}
-1      &  1     &   0    & \dots  & 0     \\
1       & -2     &   1    & \ddots & \vdots      \\
0  & \ddots & \ddots & \ddots & 0     \\
\vdots   & \ddots    & 1      & -2      & 1      \\
 0      & \dots  & 0      & 1      & -1
\end{pmatrix}
\end{equation*}

The spectrum of $\mathbf{\Delta}$ is well known \cite{Edwards}. 
The eigenvalues are simple:
\begin{equation*}
-\omega_j^2 = -4 \sin^2\left(\frac{(j-1)\pi}{2N} \right), \ j \in \{1,...,N \} .
\end{equation*}
The corresponding eigenvectors have components
\begin{equation*}
v^j_m = \sqrt{2\over N} \cos\left(\frac{(j-1)\pi}{N}(m-\frac{1}{2}) \right), ~~~ j,m \in \{1,...,N\}.
\end{equation*}

\section{Existence of 2-mode solutions}
\label{2mode}
We seek nonlinear solutions of (\ref{phi4}) that involve two nonlinear normal modes.
Substituting the ansatz,
\begin{equation}
\label{ansatz2}
\mathbf{u}(t)=a_j(t) \mathbf{v}^j +a_k(t) \mathbf{v}^k,
\end{equation}
into the equation of motion (\ref{phi4}) and projecting on each mode $\mathbf{v}^j$ and $\mathbf{v}^k$, 
we get 
\begin{equation*}
\left\{
\begin{array}{r c l}
\ddot{a}_{j}=-\omega_{j}^{2} a_{j} - \sum_{m=1}^{N} u_m^3 v_m^j , \\
\\
\ddot{a}_{k}=-\omega_{k}^{2} a_{k} - \sum_{m=1}^{N} u_m^3 v_m^k ,
\end{array}
\right.
\end{equation*}
where we have used the orthogonality of the eigenvectors 
$\prodscal{\mathbf{v}^{j}}{\mathbf{v}^{k}} =0$. 
The term $u_{m}^{3}$ can be written as 
\[ u_m^3=a_j^3 (v_m^j)^3 + 3 a_j^2 a_k (v_m^j)^2 v_m^k + 3 a_j a_k^2 v_m^j (v_m^k)^2 + a_k^3 (v_m^k)^3 , \]
so that the above equations can be written
\begin{eqnarray*}
\ddot{a}_{j}=-\omega_{j}^{2} a_{j} 
- a_j^3 \sum_{{m=1}}^{N} (v_m^j)^4 
-3 a_j^2 a_k \sum_{{m=1}}^{N} (v_m^j)^3 v_m^k \\
-3 a_j a_k^2 \sum_{{m=1}}^{N} (v_m^j)^2 (v_m^k)^2 - a_k^3 \sum_{{m=1}}^{N} (v_m^k)^3 v_m^j, \\
\ddot{a}_{k}=-\omega_{k}^{2} a_{k} 
- a_k^3 \sum_{{m=1}}^{N} (v_m^k)^4 
-3 a_j^2 a_k \sum_{{m=1}}^{N} (v_m^j)^2 (v_m^k)^2 \\
-3 a_j a_k^2 \sum_{{m=1}}^{N} v_m^j (v_m^k)^3 - a_j^3 \sum_{{m=1}}^{N} (v_m^j)^3 v_m^k .
\end{eqnarray*}

To have two periodic solutions for $a_{j}$ and $a_k$, these
equations should be uncoupled and this imposes
\begin{equation}
\label{cond2}
\sum_{{m=1}}^{N} (v_m^j)^3 v_m^k = 0, ~~
\sum_{{m=1}}^{N} (v_m^j)^2 (v_m^k)^2=0, ~~
\sum_{{m=1}}^{N} v_m^j (v_m^k)^3=0,
\end{equation}
in which case the equations reduce to
\begin{equation*}
\left\{
\begin{array}{l c r}
\ddot{a}_{j}=-\omega_{j}^{2} a_{j} - {1 \over N-S_j} a_j^3, \\
\\
\ddot{a}_{k}=-\omega_{k}^{2} a_{k} - {1 \over N-S_k} a_k^3 .
\end{array}
\right.
\end{equation*}
where $S_j$ is the number of soft nodes of $\mathbf{v}^j$ and 
$S_k$ is the number of soft nodes of $\mathbf{v}^k$.

The condition (\ref{cond2}) is the criteria for the existence of $2$-mode solutions.
It is clear that not all nonlinear modes satisfy this condition.
From the examples mentioned in section \ref{exp}, 
only cycles where $N$ is multiple of $4$ exhibit 
nonlinear normal modes satisfying the condition (\ref{cond2}), 
they are the nonlinear modes $\mathbf{v}^{N\over 2}$ and $\mathbf{v}^{{N\over 2}+1}$
corresponding to the double frequency $\omega_{N\over2} = \omega_{{N\over2}+1}=\sqrt{2}$.

\end{appendices}

\section*{Acknowledgment}

This work is part of the XTerM project, co-financed by the European Union with the European regional development fund (ERDF) and by the Normandie Regional Council.

\end{document}